\documentclass[aps, prb, twocolumn, amsmath, amssymb, floats, floatfix, longbibliography, superscriptaddress, 10pt]{revtex4-2}

\usepackage{graphicx}
\usepackage{dcolumn}
\usepackage{bm}
\usepackage{hyperref}
\usepackage{braket}
\usepackage{tikz}
\usetikzlibrary{quantikz}
\usepackage{adjustbox}
\definecolor{cadmiumgreen}{rgb}{0.0, 0.42, 0.24}

\begin{document}

\title{Diagnosing entanglement dynamics in noisy and disordered spin chains via the measurement-induced steady-state entanglement transition}



\author{T. Boorman}
\email{tjb8@st-andrews.ac.uk}
\affiliation{Department of Physics, Lancaster University, Lancaster LA1 4YB, United Kingdom}
\affiliation{School of Physics and Astronomy, University of St Andrews, St Andrews KY16 9SS, United Kingdom}

\author{M. Szyniszewski}
\affiliation{Department of Physics, Lancaster University, Lancaster LA1 4YB, United Kingdom}
\affiliation{Department of Physics and Astronomy, University College
  London, London WC1E 6BT, United Kingdom}

\author{H. Schomerus}
\affiliation{Department of Physics, Lancaster University, Lancaster LA1 4YB, United Kingdom}
\author{A. Romito}
\affiliation{Department of Physics, Lancaster University, Lancaster LA1 4YB, United Kingdom}

\begin{abstract}
We utilize the concept of a measurement-induced entanglement transition to analyze the interplay and competition of processes that generate and destroy entanglement in a one-dimensional quantum spin chain  evolving under a locally noisy and disordered Hamiltonian.
We employ continuous measurements of variable strength to induce a transition from volume to area-law scaling of the steady-state entanglement entropy.
While static background disorder systematically reduces the critical measurement strength,
this critical value depends non-monotonically on the strength of non-static noise.
According to the extracted fine-size scaling exponents, the universality class of the transition is independent of the noise and disorder strength.
We interpret the results in terms of the effect of static and non-static disorder on the intricate dynamics of the \emph{entanglement  generation rate} due to the Hamiltonian in the absence of measurement, which is fully reflected in the behavior of the critical measurement strength.
Our results establish a firm connection between this entanglement growth and the steady-state behavior of the measurement-controlled systems, which therefore can serve as a tool to quantify and investigate features of transient entanglement dynamics in complex many-body systems via a steady-state phase transition.
\end{abstract}

\maketitle

\section{Introduction}
\label{sec:Introduction}

Despite the reversible nature of unitary dynamics, closed many-body quantum systems  can exhibit the hallmarks of thermalization. This apparent paradox is explained by the eigenstate thermalization hypothesis (ETH), stating that local observables of generic many-body systems exhibit ergodic dynamics after a finite time~\cite{Deutsch1991, Srednicki1994, DAlessio2016, Borgonovi2016, garrison2018does, kim2014testing, Bao2019Eigenstate, nandkishore2014review, abanin2017recent, Abanin2019Review}. The ergodic regime supports near-maximal entanglement between different parts of the system, which results in an extensive scaling of the entanglement entropy, known as the volume law. However, this behavior is violated in systems where entanglement generation is inhibited, as in the prominent case of many-body localized (MBL) one-dimensional interacting systems featuring sufficiently strong local disorder \cite{abanin2017recent, Abanin2019Review, pal2010many, nandkishore2014review, Prosen2008Many, Basko2006, Gornyi2005, Altman2015, chandran2015constructing, luitz2015many, Parameswaran2018}. In such systems, the entanglement entropy asymptotically becomes independent of system size, a behavior that is known as an area law.

Recently, a different mechanism for departing from the volume law has garnered substantial interest,
where this is achieved in the out-of-equilibrium dynamics of open systems. This mechanism is based on local measurements occurring frequently enough, which, due to the quantum Zeno effect, can drive a measurement-induced entanglement transition (MIET) to a sub-extensive scaling of the steady-state entropy~\cite{Li2018, Chan2018, Skinner2018, Li2019, Szyniszewski2019, Zabalo2020, Napp2019, Fan2021, Gullans2019purification, Bao2020, Bera2020, Jian2020, Li2020Conformal, Szyniszewski2020universality, LopezPiqueres2020, Shtanko2020, Lavasani2021, Sang2021measurement, Zhang2020, Choi2020, Turkeshi2020, Gullans2020, Nahum2021, Cao2019, Alberton2021, Tang2020, Goto2020, Fuji2020, Rossini2020, Lunt2020, Chen2020, Liu2021, Biella2020, Gopalakrishnan2021, Jian2021, Tang2021, Turkeshi2021, Buchhold2021, Lang2020, Ippoliti2021entanglement, VanRegemortel2021, Vijay2020, Nahum2020defects, Li2021, Gullans2020, Lunt2020hybridity, Gullans2020lowdepth, Fidkowski2021, Maimbourg2021, Iaconis2020, Ippoliti2021postselection, Lavasani2020topological, Sang2020entanglement, Shi2020, Bao2021, Rossini2021}. The transition stems from the competition between the entanglement growth rate due to the unitary dynamics~\cite{tsomokos2007,knap2018,ghosh2021} and the rate of measurements localizing the state. The paradigmatic model of MIET consists of a one-dimensional random quantum circuit, where the system undertakes a stochastic evolution via the application of random two-site unitary matrices arranged in a running-bond pattern, and intercalated by local projective measurements occurring with finite probability~\cite{Li2018, Chan2018, Skinner2018, Li2019}. A number of works have characterized this transition with gates sampled either uniformly from the Haar measure of the unitary group, or from the Clifford group, which preserves computational tractability for large system sizes.  While the MIETs in Haar and Clifford circuits appear to belong to different universality classes~\cite{Zabalo2020}, the former is continuously connected to the transition in models with continuous stochastic dynamics~\cite{Szyniszewski2020universality}. Further studies have addressed explicitly unitary dynamics generated by a Hamiltonian, ranging from non-interacting free fermions~\cite{Cao2019, Alberton2021,Buchhold2021} to interacting systems~\cite{Tang2020, Goto2020, Fuji2020, Rossini2020, Lunt2020}, also in combination with post-selected measurements~\cite{Chen2020, Gopalakrishnan2021, Biella2020, Jian2021, Tang2021, Liu2021, Turkeshi2021}.

When the system is described by a Hamiltonian, the entanglement dynamics depend on physical
processes that generate or suppress entanglement, and facilitate or inhibit its propagation over the system. The measurement-induced entanglement transition should then generically reflect such dependence, and hence provide a tool to study it under steady-state conditions. So motivated, we here investigate the entanglement dynamics of a continuously measured one-dimensional chain of interacting spins subject to local time-dependent fluctuations (noise) and static background disorder. In absence of measurements, the system exhibits intricate entanglement dynamics, arising from the competition of static disorder driving the system towards MBL behavior, and temporal noise facilitating the entanglement generation and propagation. Controlling this system by continuous measurements of a variable strength enables us to interpolate between the free unitary dynamics, and the strong projective measurements commonly seen in the study of random quantum circuits. The entropy in the steady state then reflects the interplay of time scales related to the measurements, interactions, temporal noise, and static disorder in the system, each tied to a specific physical parameter in the model. This allows us to discriminate the separate effects of these processes, and develop a picture of the overall entanglement dynamics of the system.

Our analysis confirms the existence of a phase boundary between extensive and sub-extensive entanglement scaling in the parameter space spanned by the temporal noise, background disorder, and measurement strength, separating regions where a volume law or area law is obeyed.
The critical measurement strength for the entanglement transition depends non-monotonically on the temporal noise, with an initially increasing behavior followed by a decrease over a large range of noise strengths. Static disorder systematically reduces the  critical measurement strength, but not down to zero, and
increases the parameter range for which the noise facilitates the entanglement.
We further show that these features capture fine details of the transient entanglement dynamics of the closed system,
including the non-monotonous parameter dependence of entanglement growth in the limit of a noisy system without a static disorder ~\cite{knap2018}, and the maintenance of volume-law dynamics in non-stationary MBL systems~\cite{Lunt2020}.
This connection between transient and steady-state entanglement dynamics is further solidified by our observation that
the transition remains in the same universality class as found for continuous measurements of random, stochastic unitary evolution \cite{Szyniszewski2020universality},
embodied in a critical exponent of the correlation length that is independent of the noise and disorder strength.


The model along with a description of its physical parameters and implementation is introduced in Sec.~\ref{sec:Model}.
Our main results are collected in Sec.~\ref{sec:Results}.
The existence of the entanglement transition in the continuum limit of systems with noise and disorder is established in Sec.~\ref{sec:Extracting_Lambda_Section}. In Sec.~\ref{sec:noise}, we investigate how the transition depends on the noise strength, leading us to identify the
nonmonotonous dependence of the critical measurement strength, and show that this is mirrored in the transient entanglement dynamics.
In Sec.~\ref{sec:disorder}, we extend this discussion to the case of additional static disorder, resulting in a reduction of the critical measurement strength that we again link to the transient dynamics.
The finite-size scaling analysis determining the critical exponent and establishing the universality of the transition is carried out in Sec.~\ref{sec:FSS_Analysis_Section}.
Our conclusions are collected in Sec.~\ref{sec:Conclusions}.

\section{Model and implementation}
\label{sec:Model}

\begin{figure}[t]
  \centering
  \includegraphics[width = 0.99\columnwidth]{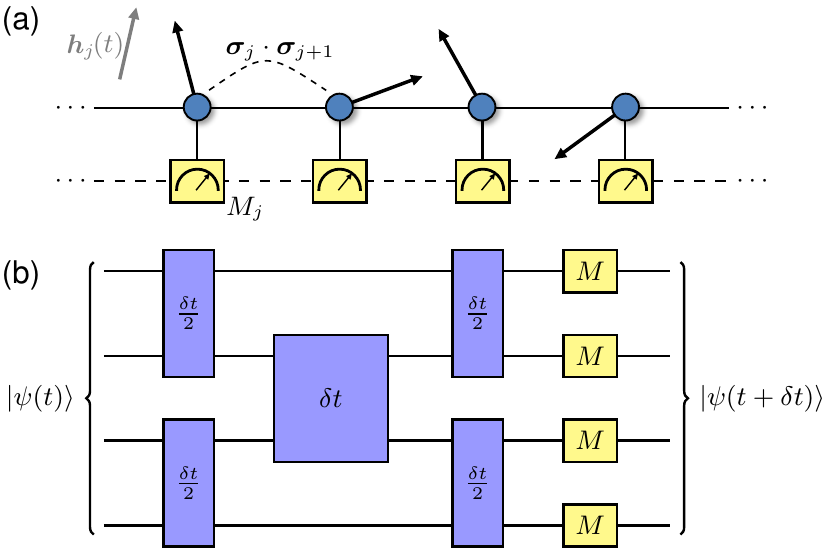}
  \caption{Model system and implementation of the time evolution. (a)~The system consists of a spin-half Heisenberg chain (blue dots) in a disordered time-dependent external field $\bm{h}_{j}$, as given by Hamiltonian~(\ref{eq:System_Hamiltonian}), where each spin is furthermore  continuously coupled to a local quantum meter $M_j$ (yellow boxes). (b)~Quantum circuit implementation of the evolution for a time step $\delta{t}$, accurate to orders $\mathcal{O}(\delta t^3)$. The blue boxes represent unitary evolution under a nearest-neighbor interaction term and the fluctuating disordered field in the Hamiltonian, broken up into layers involving sets of mutually commuting operators, while the yellow boxes implement the measurements as detailed in the text.
  }
  \label{fig:QC_Diagram}
\end{figure}

\subsection{Model}

Our model consists of a one-dimensional chain of spins with size 1/2, evolving under the combined action of unitary dynamics and continuous local measurements. The unitary evolution is generated by isotropic Heisenberg couplings between nearest-neighbor spins subject to a locally fluctuating magnetic field [cf.~Fig.~\ref{fig:QC_Diagram}(a)], as described by the Hamiltonian
\begin{equation}
  \label{eq:System_Hamiltonian}
  \hat{H}_{t} = J \sum_{j=1}^{L}\bm{\sigma}_{j}\cdot\bm{\sigma}_{j+1} + \sum_{j=1}^{L} \bm{h}_{j}(t) \cdot\bm{\sigma}_{j}.
\end{equation}
Here $L$ is the number of spins and $\bm{\sigma}_{j}$ is the three-component vector of Pauli matrices $\sigma_{j,\alpha}$, $\alpha=x,y,z$, where we set $\bm{\sigma}_{L+1} \equiv \bm{\sigma}_{1}$ to impose periodic boundary conditions. The magnetic field, $\bm{h}_{j}(t)$, combines spatially uncorrelated white-noise temporal fluctuations with static background disorder, as incorporated in the time averages
\begin{align}
  & \overline{h_{j,\alpha}(t)} = h_{j,\alpha} , \\
  & \overline{\delta h_{j,\alpha}(t) \delta h_{k,\beta}(t)}= J \hbar \, \xi_r^2 \delta (t-t') \delta_{j,k} \delta_{\alpha,\beta},
\end{align}
where $\delta h_{j,\alpha}(t) = h_{j,\alpha}(t) - h_{j,\alpha}$.
Note that the time-averaged background field $h_{j,\alpha}$ is still spatially dependent. We assume this static background to be disordered, with no spatial correlations, which we implement by drawing $h_{j,\alpha}$ independently for different $j$ and $\alpha$ from a uniform distribution over an interval $[-\xi_s J, \xi_s J]$.
As the parameter $J>0$ sets the energy scale of the interactions that generate the entanglement, while $\hbar/J$ sets the associated time scale, we use these as the fundamental units of the problem, which is equivalent to setting $J = \hbar = 1$. The unitary dynamics is therefore controlled by the  two dimensionless parameters $\xi_r$ and $\xi_s$, which characterize, respectively, the strength of the temporal noise $\delta h_{j,\alpha}(t)$ and spatial background disorder $h_{j,\alpha}$, relative to the deterministic energy scale $J$. The ability of controlling  these two parameters independently allows us to address separately the effect of  static disorder driving the system towards a many-body localized state, and delocalizing noise
~\cite{pal2010many, nandkishore2014review, Prosen2008Many, abanin2017recent, Abanin2019Review}.

The measurement process consists of independent continuous Gaussian measurements \cite{Jacobs2014quantum, wiseman_milburn_2009} of the $Z$\nobreakdash-component of each spin in the chain. For the $j$\nobreakdash-th spin, the effect of the measurement is described by the Wiener process
\begin{equation}
\label{eq:Wiener}
  \ket{\psi} \to \mathcal{N} \left[ 1  - \sum_j \left( \lambda^2 \delta t \, \langle \sigma_{j,z} \rangle + \delta W_j \right) \sigma_{j,z} \right] \ket{\psi},
\end{equation}
where $\delta t=J dt/\hbar$ is a dimensionless infinitesimal time interval, the random variables $W_j$ are independently Gaussian-distributed with zero mean and variance $\lambda^2 \delta t$, and $\mathcal{N}$ is a normalization constant. The measurement process is fully determined by the dimensionless constant  $\lambda^2$, which sets the inverse measurement time, so that for time $t \gg \hbar/J \lambda^2$ an isolated continuously measured spin localizes at one of the eigenvalues of $\sigma_{j,z}$. Thereby, the continuous measurement-induced dynamics are obtained from a sequence of sufficiently frequent measurements with controllable system-detector coupling.

\subsection{Implementation of the time evolution}

Numerically, we implement the continuous-time evolution of the system under the combined effect of measurements and unitary evolution including the stochastic noise by approximating the limiting process of an infinitesimal discrete step $\delta t$. 
This implementation is illustrated in Fig.~\ref{fig:QC_Diagram}(b). For each time step, we draw the values of $\delta h_{j,\alpha}(t)$ from a uniform distribution $[-\xi_r / \sqrt{\delta t},\xi_r / \sqrt{\delta t}]$. The scaling with the time-interval $\delta t$ allows us to recover the white-noise correlations as $\delta t \to 0$. In analogy with the running-bond pattern of common quantum circuits, we arrange the summands of Hamiltonian~(\ref{eq:System_Hamiltonian}) to form two sets consisting of mutually commuting operators. From here we employ the symmetrized second-order Suzuki-Trotter expansion~\cite{hatano2005finding, christof98Simulating}. This reduces unitary evolution from multiplication by a $2^{L} \times 2^{L}$ matrix to the action of a series of $4\times4$ matrices, whilst preserving accuracy up to $\mathcal{O}(\delta{t}^3)$. The full unitary evolution of the system is obtained by applying successive layers of this elementary unitary block.

\begin{figure}[b]
    \centering
    \includegraphics[width = 0.99\columnwidth]{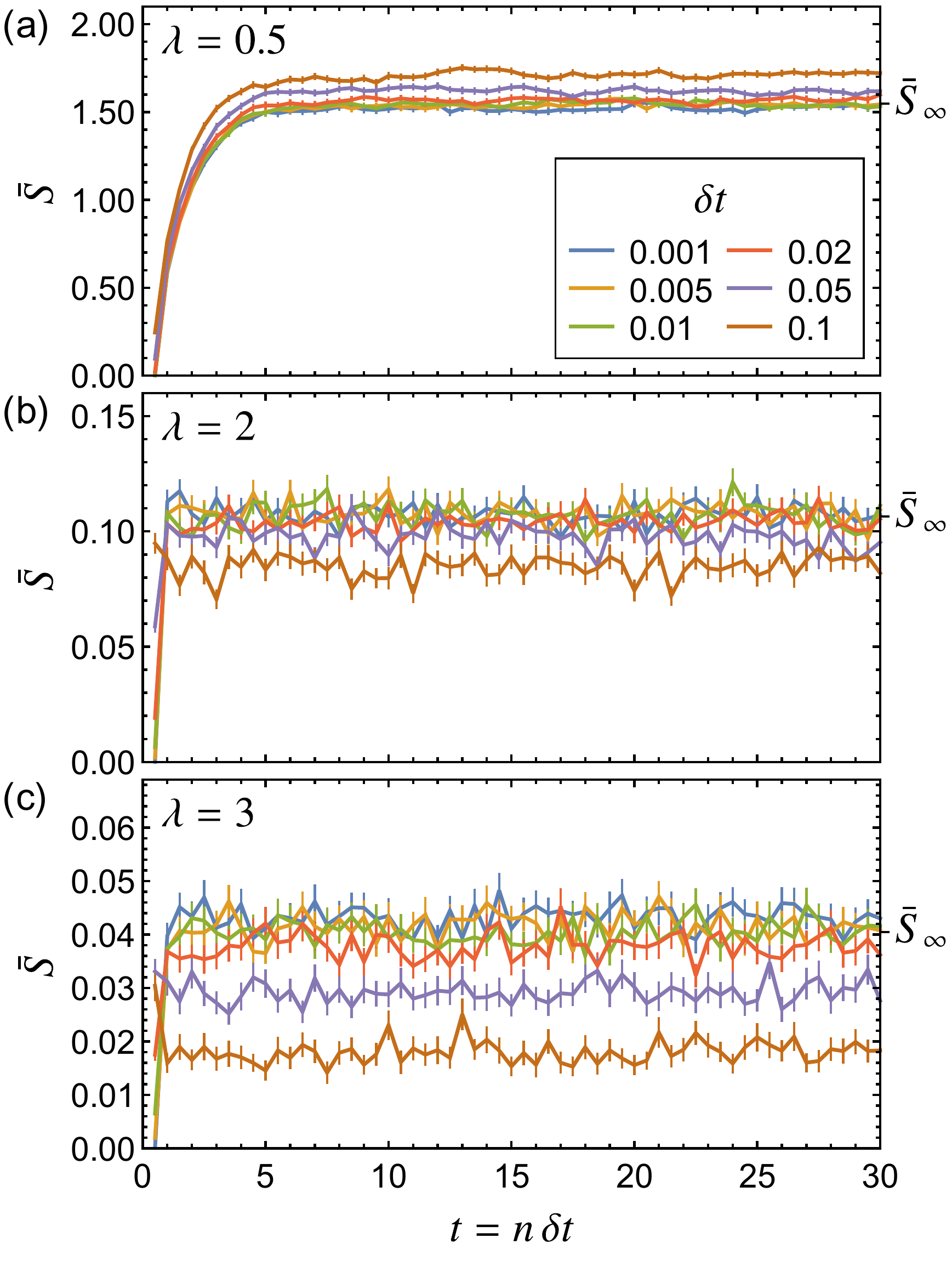}
    \caption{Approach of the continuum limit of the entanglement dynamics by reduction of the discretization time step $\delta t$.
    The panels show the realization-averaged half-chain entanglement entropy, $\bar{S}$, for  measurement strengths (a)~$\lambda = 0.5$, (b)~$\lambda = 2$, and (c)~$\lambda = 3$,  and values of $\delta t$ as specified in the figure. The system size is fixed to $L=12$, and the noise and disorder strengths are set to $\xi_r = 2$, $\xi_s=0$, while the initial state is set to a separable N\'eel state. The curves are then obtained by averaging over 500 realizations. They collapse for $\delta t \lesssim 0.01$, identifying this as a suitable value to faithfully approximate the continuous-time evolution. We use such simulations to infer both the transient dynamics of the system out of the initial state, as well as the steady-state value $\bar{S}_\infty$ approached at long times.}
    \label{fig:continuumlim}
\end{figure}

Between each layer of unitary evolution, we perform independent measurement operations $M$ sampling the possible outcomes at each site. For a pre-measurement state $\ket{\psi}$, the measurement outcome $x \in \mathbb{R}$ for a given spin is drawn from the probability distribution
\begin{equation}
  P(x) = \bra{\psi} M^\dagger_x M_x \ket{\psi},
  \label{eq:x_distribution}
\end{equation}
where $M_x = \phi_{p}(x-\lambda_0) \hat{\Pi}_{+}^{(j)} + \phi_{p}(x+\lambda_0) \hat{\Pi}_{-}^{(j)}$ is the Kraus operator associated with this outcome, with $\phi_p(x) = \pi^{-1/4}\exp(-{x}^{2} / {2})$, and  $\hat{\Pi}_{\pm}^{(j)}$ the  projectors onto spin up and down.
Conditional to the specific outcome $x$ obtained, the post-measurement state is given by
\begin{equation}
  \ket{\psi'} = \frac{1}{\sqrt{P(x)}} M_x \ket{\psi}.
  \label{eq:Measurement_Backaction}
\end{equation}
These discrete measurements are characterized by the parameter $\lambda_0 = \sqrt{\delta t} \lambda$, where this scaling reproduces the stochastic continuous dynamics in Eq.~(\ref{eq:Wiener}) for $\delta t\to 0$. The variable continuous-measurement strength $\lambda$ then gives us the ability to interpolate between the cases of no measurements, $\lambda = 0$, and strong projective measurements, $\lambda \to \infty$.

\subsection{Verification of the continuum limit}

The main quantity of interest in this work is the entanglement entropy of the system. Specifically, we consider the bipartite von-Neumann entanglement entropy of a sub-system $A$, which is defined as
\begin{equation}
\label{eq:vN}
    S_{A}(t) = -\textrm{tr}\left[ \rho_{A}(t) \ln \rho_{A}(t)\right],
\end{equation}
where $\rho_A (t)$ is the reduced density matrix of a subsystem $A$, obtained by tracing the system's density matrix $\rho$ over the subsystem complementary to $A$.

Before analyzing its behavior in detail, we first establish that this quantity can be  faithfully  obtained in the described numerical implementation of the combined stochastic process arising from the temporal fluctuations and continuous measurements.
This is illustrated in  Fig.~\ref{fig:continuumlim}, which shows the time-dependence of the realization-averaged entanglement entropy $\bar{S}$ at a fixed value of $\xi_r=2$, and for three different values of the measurement strength $\lambda$, while $\xi_s=0$. The initial state is a separable N\'eel state, $\ket{\psi} = \ket{\uparrow \downarrow \uparrow \downarrow \cdots}$, which we will maintain throughout the study as the steady-state distribution is consistently observed to be independent of the initial state. We also note that since $S_A$ is a non-linear function of $\rho_A$, a crucial feature of the described measurement-induced dynamics is that it always keeps the  system in a  pure state, so that the entanglement entropy can be computed before averaging.

As seen in the figure, the average entanglement entropy increases with time until it saturates around a steady-state value, $\bar{S}_\infty$, that depends on the measurement strength $\lambda$. The consistent behavior of the dynamics for small values of $\delta t$ demonstrates the validity of our discrete time-evolution model to simulate the continuum dynamics of the system.
We verified that
a similarly consistent behavior is obtained in systems with finite background disorder $\xi_s$.
Examining such results, we found $\delta{t} = 0.01$ to be sufficiently small to faithfully approximate the continuous dynamics over the range of values for $\xi_r$, $\lambda$, and $\xi_s$ considered in this paper.

The results in Fig.~\ref{fig:continuumlim} already illustrate how measurements reduce the steady-state entanglement entropy  $\bar{S}_\infty$ in the system.
Having identified the continuum limit, we next study these features of the entanglement dynamics in detail.


\section{Results}
\label{sec:Results}

We now turn to the detailed analysis of the entanglement dynamics, where we establish the existence of a steady-state entanglement transition and examine its relation to the transient entanglement dynamics, and first introduce the key quantities involved in these considerations.

In the steady state, the entanglement entropy ~\eqref{eq:vN} is expected to exhibit extensive (volume-law) scaling $\bar{S}_\infty\propto L$ in the ergodic regime,
and sub-extensive scaling in the regime dominated by measurements.
In particular, for a fully ergodic system, in which the system's state is distributed extensively over all available states in the Hilbert space, the steady-state half-chain entropy assumes a known form predicted by Page~\cite{Page93Avg},
\begin{equation}
    \bar{S}_\infty^{\mathrm{erg}} = \frac{L}{2}\textrm{ln}(2) - \frac{1}{2}.
    \label{eq:Page_curve}
\end{equation}
In the following, we analyze the steady-state behavior by computing the average $\bar{S}_\infty$ over different realizations of the stochastic dynamics, as well as the corresponding cross-realization variance (CRV) $\textrm{var} \, S_\infty$. 
The steady-state average entropy, $\bar{S}_\infty$, is determined by averaging over realizations of a given noise and measurement strength, with the contribution from each realization being found by averaging over $100$ points within the steady regime, uniformly spaced according to $100 \delta{t} = 1$.
Throughout, we take $L$ to be even and $A$ to be half of the chain.
The dependence of $\bar{S}_\infty$ on the system size distinguishes the volume law from sub-volume law scaling, while $\textrm{var}\, S_\infty$ is expected to reveal large sample-to-sample fluctuations near the transition. Both features can drift with system size, which we address by finite-size scaling.
For systems with length $L\in 4\mathbb{N}$, we also employ the tripartite mutual information (TMI),  defined as
\begin{align}
    \mathcal{I}_3 (A:B:C) &= S_{A} + S_{B} + S_{C} + S_{A\cup B \cup C} \nonumber \\
    &- S_{A\cup B} - S_{A\cup C} - S_{B\cup C},
\end{align}
where $A$, $B$ and $C$ are adjacent subsystems of length $L/4$.
The averaged TMI is expected to vanish asymptotically with increasing system size in systems with an area law, to take negative values proportional to $L$ for a volume law, and to approach a finite negative value at the critical point~\cite{Zabalo2020}. In stroboscopic (discrete-time) models, $\mathcal{I}_3$ was shown to exhibit minimal finite-size effects at the critical point~\cite{Zabalo2020}, which could then be located by finding the crossing points of $\mathcal{I}_3$ for varying system sizes. However, in continuous-time models, crossings of $\mathcal{I}_3$ have been found to drift slightly with increasing system size~\cite{Szyniszewski2020universality}, implying the need for further finite-size analysis also for this quantity.


\begin{figure}[t!]
    \centering
    \includegraphics[width = 0.99\columnwidth]{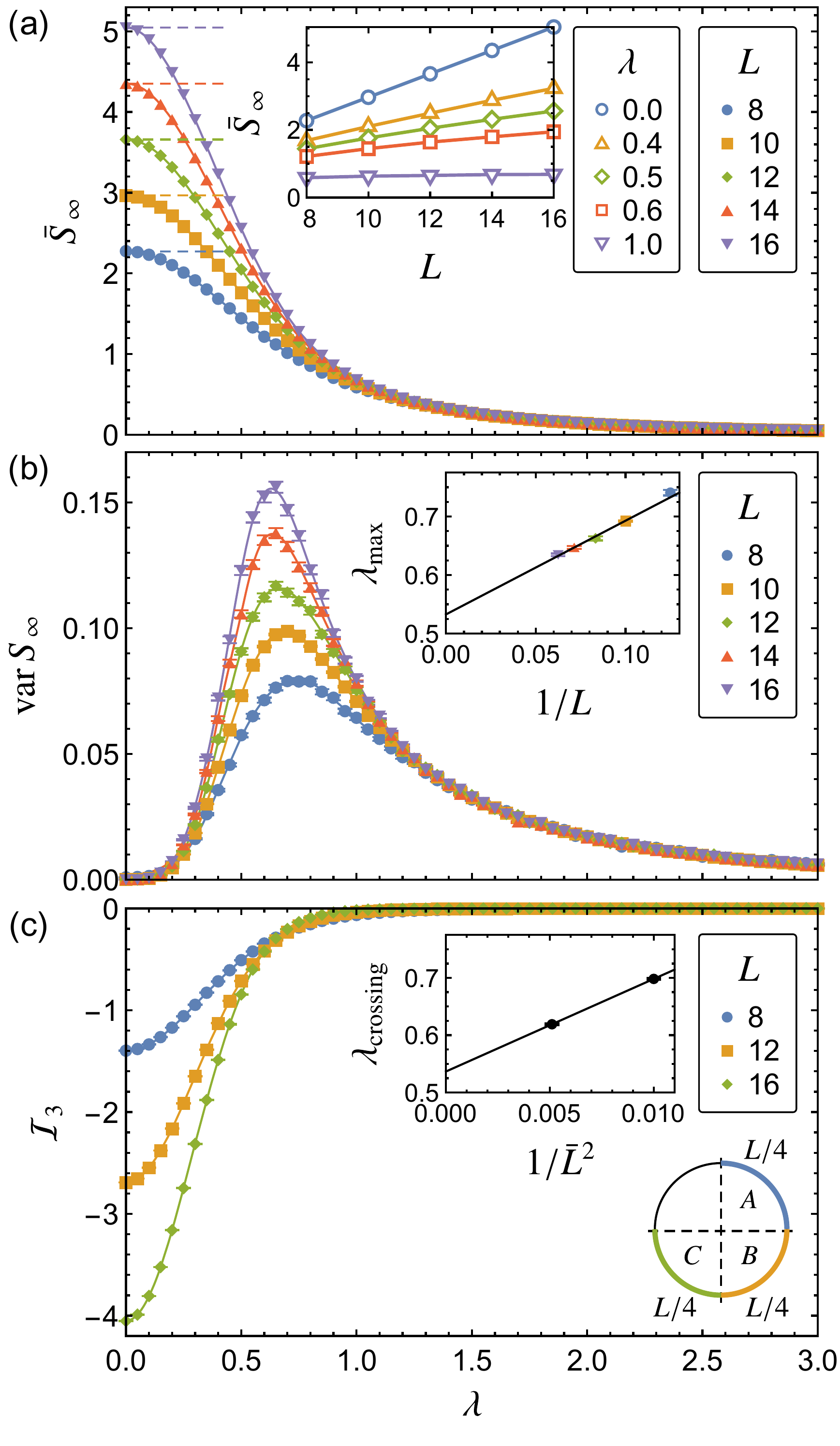}
    \caption{Steady-state entanglement transition in a noisy system ($\xi_r = 1.3$) without static background disorder ($\xi_s=0$), induced by varying the measurement strength $\lambda$.
    Panel (a) shows the steady-state entanglement entropy $\bar{S}_\infty(\lambda)$ for specified system sizes $L$, obtained by averaging over 100 realizations. This approaches the extensive Page law~(\ref{eq:Page_curve}) for $\lambda\to 0$, indicated by the dashed lines,  and becomes independent of system size when $\lambda$ is large.  The inset illustrates this transition in terms of the scaling of  $\bar{S}_\infty(\lambda)$ with $L$ at fixed $\lambda$.
    Panel (b)~shows the corresponding variance $\textrm{var} \, S_\infty$, which displays a maximum near the entanglement transition. The inset shows the extrapolation of this maximum to the thermodynamic limit, yielding $\lambda_c = 0.533(5)$.
    Panel (c)~shows the average tripartite mutual information $\mathcal{I}_3$, where curves for different system sizes cross near the transition point. In the inset, these crossings are extrapolated to the thermodynamic limit, where $\bar{L}$ is the average of $L$ for the crossing curves, yielding the compatible estimate $\lambda_c = 0.537(4)$.}
    \label{fig:jointplot1}
\end{figure}

\begin{figure}[t!]
    \centering
    \includegraphics[width = 0.99\columnwidth]{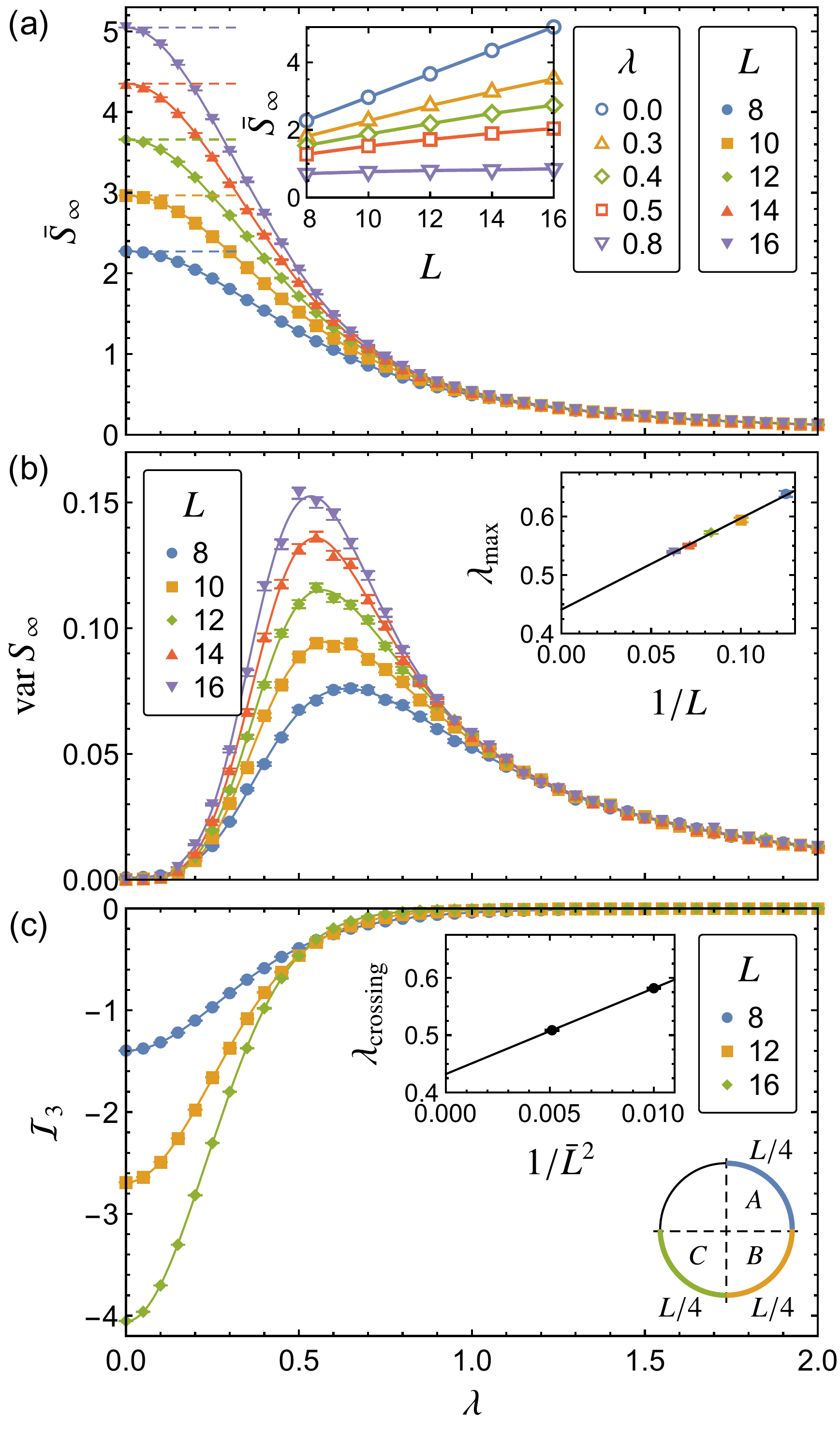}
    \caption{Analogous to Fig.~\ref{fig:jointplot1}, but in presence of additional static background disorder ($\xi_s=3$).
    This results in a reduction of the critical measurement strength, for which we obtain $\lambda_c = 0.441(5)$ by extrapolation of the location of maximal variance, and  $\lambda_c = 0.4323(31)$ by extrapolation of the crossings of the
tripartite mutual information.}
    \label{fig:jointplot2}
\end{figure}

\subsection{Entanglement dynamics and transition at fixed noise and disorder}
\label{sec:Extracting_Lambda_Section}

First, we establish the existence of a measurement-induced entanglement transition in the model when  the parameters $\xi_r$ and $\xi_s$ quantifying the noise and disorder in the unitary evolution are fixed.
Fig.~\ref{fig:jointplot1}(a) shows the dependence of $\bar{S}_\infty$ on the measurement strength $\lambda$ for systems of different sizes and a given stochastic unitary evolution ($\xi_r=1.3$) in absence of static background disorder ($\xi_s=0$). In the limit of no measurements ($\lambda=0$), we recover the expected extensive entropy of an ergodic system, given by Eq.~(\ref{eq:Page_curve}). Upon increasing the measurement strength, the entropy monotonically decreases and becomes independent of the system size. The inset shows the change from a linear scaling with system size (volume law) at $\lambda=0$  to a size-independent entropy (area law) at $\lambda=1$.

To identify the critical measurement strength $\lambda_c$ for this entanglement transition in the thermodynamic limit, we employ two different methods: extrapolation of the location of the peak in the  cross-realization variance $\textrm{var}\,S_\infty$, and extrapolation of the crossing of the tripartite mutual information for different $L$.
The cross-realization variance $\textrm{var}\,S_\infty$ is shown in Fig.~\ref{fig:jointplot1}(b).
To estimate the argument $\lambda_\textrm{max}$ where the variance is maximal, we fit the data points to an ansatz $\textrm{var}\,S_\infty\sim\exp(\sum_i a_i \lambda^i)$, indicated by the solid curves. The number of terms in the sum is determined by the reduced $\chi^2$ and the resulting best-fit parameters and covariance matrix are used in bootstrap Monte Carlo to extract $\lambda_\textrm{max}$ and its statistical error.  
Finally, assuming a linear dependence in $1/L$ we extrapolate $\lambda_\textrm{max}$ to infinite system size (cf.~inset), and obtain $\lambda_c = 0.533(5)$.

This result is corroborated by TMI results reported in Fig.~\ref{fig:jointplot1}(c). To determine the crossings of curves with different $L$, we first calculate the difference in data points between adjacent values of $L$, and then fit the results to a form $\mathcal{I}_3\sim-\exp(\sum_i a_i \lambda^i)$ (solid curves in the figure). Bootstrap Monte Carlo is then used to extract the crossing points, which are subsequently used in the extrapolation to infinite system size (see the inset) assuming a linear dependence in $\bar{L}^{-2}$~\cite{Szyniszewski2020universality}, where $\bar{L}$ is the average of $L$ for the two curves. While this extrapolation is based on only two points,  the resulting value $\lambda_c = 0.537(4)$ is in excellent agreement with the stated value from the extrapolation of the argument $\lambda_\textrm{max}$ of maximal $\textrm{var}\,S_\infty$.

Figure 4 shows the analogous results for a system with additional static background order, set to a strength $\xi_s=3$. This results in the same qualitative behavior, supporting again the existence of an entanglement transition, but with a reduced critical measurement strength. Extrapolating the results to the thermodynamics limit, we now find $\lambda_c = 0.441(5)$ from the location of maximal variance, and  $\lambda_c = 0.4323(31)$ from the crossings of the
tripartite mutual information. We next study these trends systematically, and relate them to the transient entanglement dynamics of the system.

\subsection{Dependence of the entanglement transition on noise}
\label{sec:noise}

\begin{figure}[t!]
    \centering
    \includegraphics[width=\columnwidth]{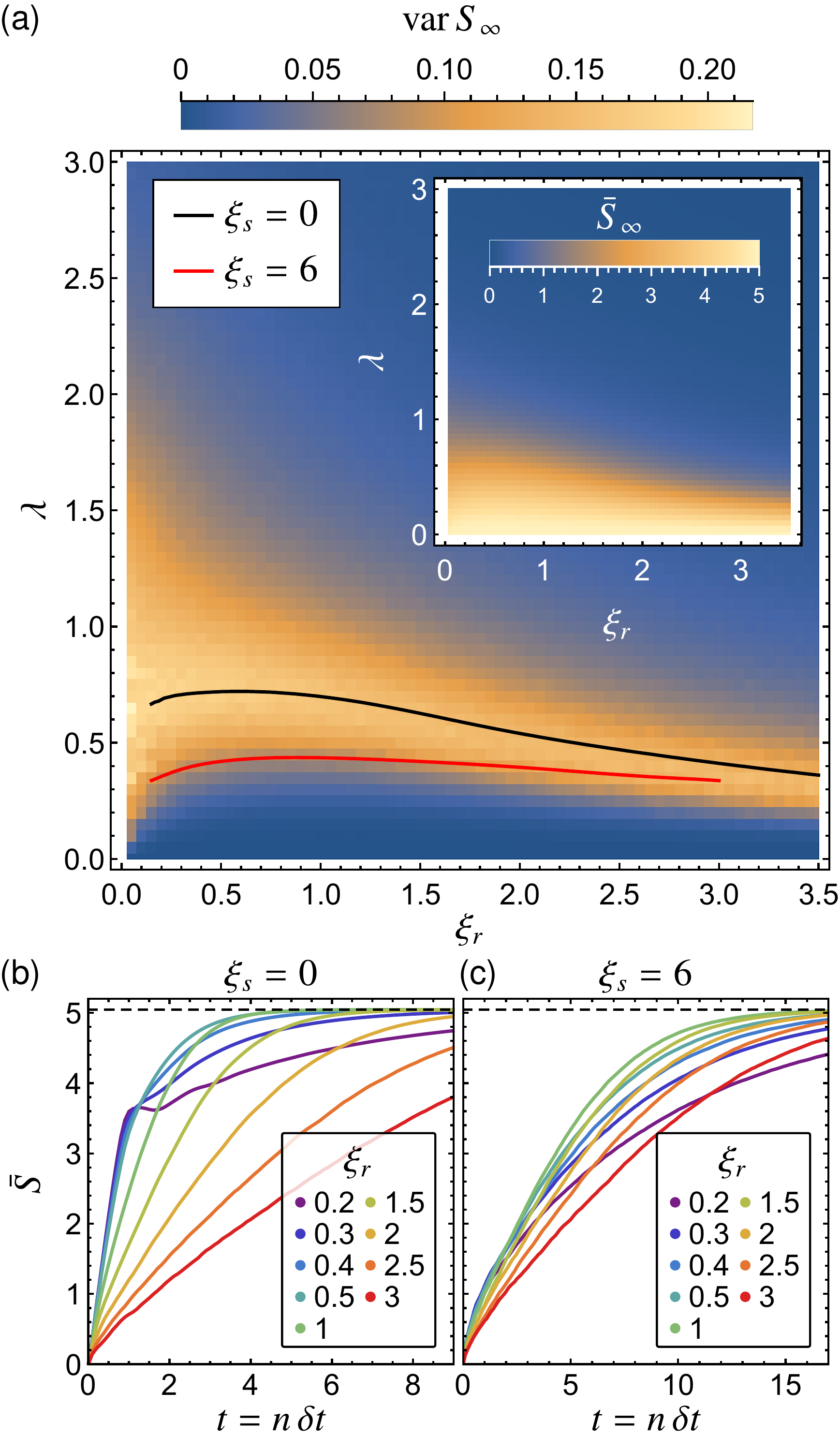}
    \caption{Dependence of the entanglement transition on the noise strength $\xi_r$. Panel (a) shows the color-coded variance $\textrm{var}\,S_\infty$ as a function of $\xi_r$ and the measurement strength $\lambda$, with $\xi_s=0$ and $L=16$. Each point represents data from 100 realizations. The black curve marks the location of maximal variance, which serves as an estimate of the critical measurement strength $\lambda_c$. This separates regions of small and large steady-state entanglement entropy $\bar{S}_\infty$, as illustrated in the inset. The red curve shows the  corresponding phase boundary for additional static background disorder $\xi_s=6$. Both phase boundaries display a nonmonotonous dependence on the noise strength $\xi_r$, which also manifests itself in the transient entanglement dynamics $\bar S(t)$ out of the separable initial state, as shown for $\xi_s=0$ in panel (b) and $\xi_s=6$ in panel (c).
     For large times, the entropy approaches the Page limit \eqref{eq:Page_curve}, indicated by the dashed lines. Weak disorder increases the rate by which this limit is approached, while strong noise reduces it, with the maximal rate obtained at $\xi_r \approx 0.5$ in panel (b) and  $\xi_r\approx 1$ in panel (c).}
    \label{fig:L16_Maps}
\end{figure}

We now examine the detailed dependence of the entanglement transition on the strength of temporal and spatial fluctuations. For this, we apply the same procedure as in the previous Sec.~\ref{sec:Extracting_Lambda_Section} to systems with different parameters combinations $\xi_r$ and $\xi_s$, and collect the results in a phase diagram, where we combine information from the average $\bar{S}_\infty$ and cross-realization variance $\textrm{var}\,S_\infty$ of the steady-state entanglement entropy.

As above, we first consider the case of finite temporal fluctuations $\xi_r$ at vanishing  spatial background disorder ($\xi_s = 0$). The results for  a system of fixed
size $L = 16$  are reported in Fig.~\ref{fig:L16_Maps}. Panel (a) shows the color-coded averaged entropy $\bar{S}_\infty$ (inset) and cross-realization variance $\textrm{var}\,S_\infty$ (main panel) as a function of the noise strength $\xi_r$ and measurement strength $\lambda$.
At any finite $\xi_r$, the averaged entropy $\bar{S}_\infty$   decreases monotonically upon increasing the measurement strength, and consistently  approaches  the value Eq.~(\ref{eq:Page_curve}) in the limit of vanishing measurements.
The main panel displays a clearly visible ridge of comparatively high variance that separates the regions where $\bar{S}_\infty$ is large or small, delineating a phase boundary between ergodic and non-ergodic dynamics. We highlight this boundary by the solid black curve, which traces out the location of maximal CRV, and take this to approximate the critical measurement strength $\lambda_c$.

This phase boundary displays two striking trends. Firstly, the critical measurement strength  $\lambda_c(\xi_r)$ depends non-monotonically on $\xi_r$. Secondly, for increasing $\xi_r$, judged by the visibly bright region, the width of the CRV curve systematically narrows. As we show later in Sec.~\ref{sec:FSS_Analysis_Section}, these features persist in the thermodynamic limit  for large values of $\xi_s$. We next relate the first trend to the transient entanglement dynamics of the system, and later employ finite-size scaling to show that, despite the second trend, the critical fluctuations are independent of $\xi_r$, and hence remain universal.

To obtain insight into the nonmonotonous  dependence in  $\lambda_{\mathrm{max}}(\xi_r)$,
we examine the effect of temporal fluctuations on the transient entanglement dynamics of the unitary system without measurements ($\lambda=0$). As shown in Fig.~\ref{fig:L16_Maps}(b), weak noise in the range $\xi_r \lesssim 0.5$ leads to a slight increase of the entanglement growth rate from the initial unentangled state. Beyond this range, this trend is reversed, so that increasing $\xi_r$ further results in a noticeable suppression of the  entanglement generation.

This behavior finds a natural phenomenological explanation in the interplay of the processes involved in the transient entanglement dynamics. The entanglement itself is generated by the spin-spin interactions but, in isolation, these interactions describe an integrable system in which the time evolution is constrained, e.g., by the rotational invariance of the Hamiltonian.
A small amount of noise $\xi_r$ is sufficient to break these symmetries, allowing the system to access a larger portion of the Hilbert space and facilitating the  evolution into an ergodic steady state.  On the other hand, when the noise becomes too large, it drives the collective dynamics of the adjacent spins towards a nonresonant regime, and thereby hinders the effectiveness of the spin-spin interaction in creating entanglement. In the entanglement transition, these detailed mechanisms are then captured via the introduction of an additional time scale arising from the measurements.


\subsection{Effect of additional background disorder}
\label{sec:disorder}
We now further examine these trends in presence of additional static background disorder, characterized by a finite value $\xi_s$.

As a first step, we reexamine the results of the previous subsection with this strength now fixed to $\xi_s=6$. We focus on the phase boundary obtained by the ridge of maximal CRV, which is shown in the main panel of Fig.~\ref{fig:L16_Maps}(a) as the red curve. We see that the static disorder systematically reduces the critical measurement strength $\lambda_c$, which maintains its non-monotonous dependence on $\xi_r$, but with the maximum shifted towards larger values of the noise strength, a trend that survives in the thermodynamic limit [cf. Fig.~\ref{fig:critical_exponent}(c)].
As shown in Fig.~\ref{fig:L16_Maps}(c), these trends again reflect the
transient entanglement dynamics.  For all values of $\xi_r$, the given static disorder inhibits the entanglement generation compared to the case in panel (b), where this disorder is absent [note the differing time ranges over which the entanglement dynamics are displayed in these panels]. However, with the background disorder fixed to the given value, increasing the temporal noise strength $\xi_r$ now comparatively enhances the entanglement generation  up to $\xi_r \simeq 1$.
These observations are again consistent with the phenomenology of the unitary entanglement dynamics, where the static background disorder introduces signatures of many-body localization, while the temporal noise counteracts the emergent local integrability of such systems.

 \begin{figure}[th!]
    \centering
    \includegraphics[width = \columnwidth]{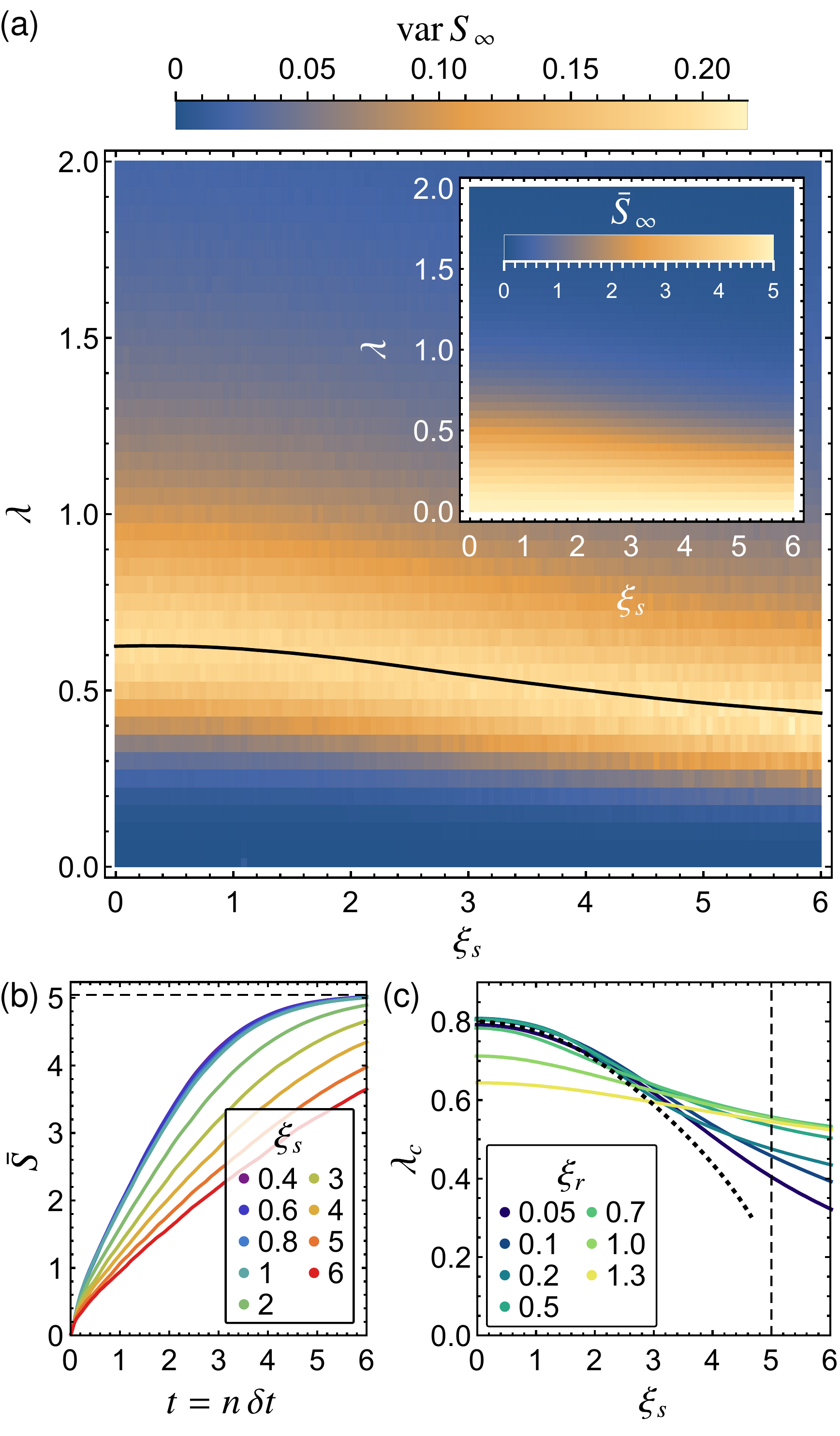}
    \caption{Dependence of the entanglement transition on the strength $\xi_s$ of static background disorder. (a)~Variance $\textrm{var}\,S_\infty$ (main panel) and average $\bar{S}_\infty$ (inset) for $L=16$ in analogy to
    Fig.~ \ref{fig:L16_Maps}(a), but in the space of disorder strength $\xi_s$ and measurement strength $\lambda$ at a fixed noise strength $\xi_r = 1.3$. The black curve again marks the location of maximal variance, resulting in an estimate of the critical measurement strength $\lambda_c$ that decreases noticeably for $\xi_s\gtrsim 1$.
    As shown in panel (b), this behavior is again reflected in the transient entanglement dynamics $\bar{S}(t)$, obtained in analogy to
    Fig.~ \ref{fig:L16_Maps}(b,c) with $\xi_s$ fixed to the indicated values and $\xi_r = 1.3$. The  averaged entanglement dynamics is nearly independent of the disorder strength as long as $\xi_s\lesssim 1$, but systematically slows down when $\xi_s$ is further increased.
    Panel (c) compares the phase boundary $\lambda_c(\xi_s)$  for various values of $\xi_r$, obtained in analogy to panel (a) but in systems of size $L=12$. The dashed vertical line marks the estimate of the MBL transition point $\xi_s^\text{crit}$ given in the literature~\cite{Geraedts2016}. The dotted line is a mere guide to the eye, indicating a possible behavior for $\xi_r\to 0$.}
    \label{fig:XsMap}
\end{figure}

To further develop this picture, we next examine the detailed dependence of the transition on the spatial disorder strength $\xi_s$ while keeping the noise strength $\xi_r$ fixed.
The overall effect of the spatial disorder on the entanglement transition is reported in Fig.~\ref{fig:XsMap}. Panel (a) shows the CRV map in the $\lambda$-$\xi_s$ space for a fixed noise strength $\xi_r=1.3$, while the inset shows the averaged steady-state entropy $\bar{S}_\infty$. In the limit $\lambda\to 0$ of vanishing measurement strength, $\bar{S}_\infty$ again tends to the entropy (\ref{eq:Page_curve}) of an ergodic system. For finite  $\lambda$, the entropy monotonously decreases with an increasing disorder strength $\xi_s$. As before, for any $\xi_s$ the entropy decreases with increasing $\lambda$, leading to two regions with relative large and small entropy.  The phase boundary is again indicated by the ridge of maximal CRV in the main panel, highlighted by the black curve and taken to approximate the critical measurement strength $\lambda_c$.
We note that as a function of increasing disorder strength, $\lambda_c(\xi_s)$ starts out flat and then decreases monotonously.

These results again agree with the phenomenological expectation that sufficiently strong spatial disorder tends to localize the system,  inducing a trend towards area-law entanglement scaling that is then attained at a comparatively reduced critical measurement strength.
This is further confirmed by examining how the transient entanglement dynamics is impacted by the static disorder. As shown in Fig.~\ref{fig:XsMap}(b), for small values of $\xi_s\lesssim 1$, the disorder does not have any noticeable impact on  the entanglement growth, while at larger values of $\xi_s$ the disorder systematically inhibits the entanglement growth.

To consolidate the results, we examine the behavior of these trends at different noise strengths $\xi_r$. For each case, we determine the CRV phase boundary as above, and collect the results in Fig.~\ref{fig:XsMap}(c).
As $\xi_r$ is reduced, the dependence of $\lambda_c$ on $\xi_s$ becomes more pronounced.
This raises the question about the specifics of the measurement-induced transition in the limit of vanishing noise, $\xi_r\to 0$. It is by now a well-documented fact that, in the absence of both noise and measurements, there exists a critical value $\xi_s^\text{crit}$ beyond which a transition to the many-body localized phase occurs~\cite{pal2010many, nandkishore2014review, Prosen2008Many, abanin2017recent, Abanin2019Review}.
Consistent with this, we see that the critical measurement strength flattens near this MBL transition point, which is estimated to be near $\xi_s^\text{crit} = 5$ in our model~\cite{Geraedts2016}.
Below this value, we observe convergence to a limiting curve that is indicated, as a rough guide to the eye, by the thick dotted line. We note that examining the entanglement transition at identically vanishing noise, $\xi_r=0$, requires caution because of the logarithmic entanglement growth towards an eventual volume state in the generic dynamics of these systems \cite{kjaell2014}. In this case, the entanglement transition takes place at a nominally vanishing measurement strength, $\lambda_c=0$, while at any finite value of $\xi_r$, $\lambda_c$ is expected to remain finite in the MBL phase \cite{Lunt2020}, indicating limits do not interchange.
This is compatible with our observations, where we find $\lambda_c$ to be finite even at very large disorder strengths. This still leaves the possibility that the
extrapolated critical measurement strength $\lim_{\xi_r\to 0}\lambda_c(\xi_s)$  develops a nonanalytic feature at the MBL transition, such as a kink.

\subsection{Thermodynamic limit and finite-size scaling analysis \label{sec:FSS_Analysis_Section}}

\begin{figure}[t!]
  \centering
  \includegraphics[width = \columnwidth]{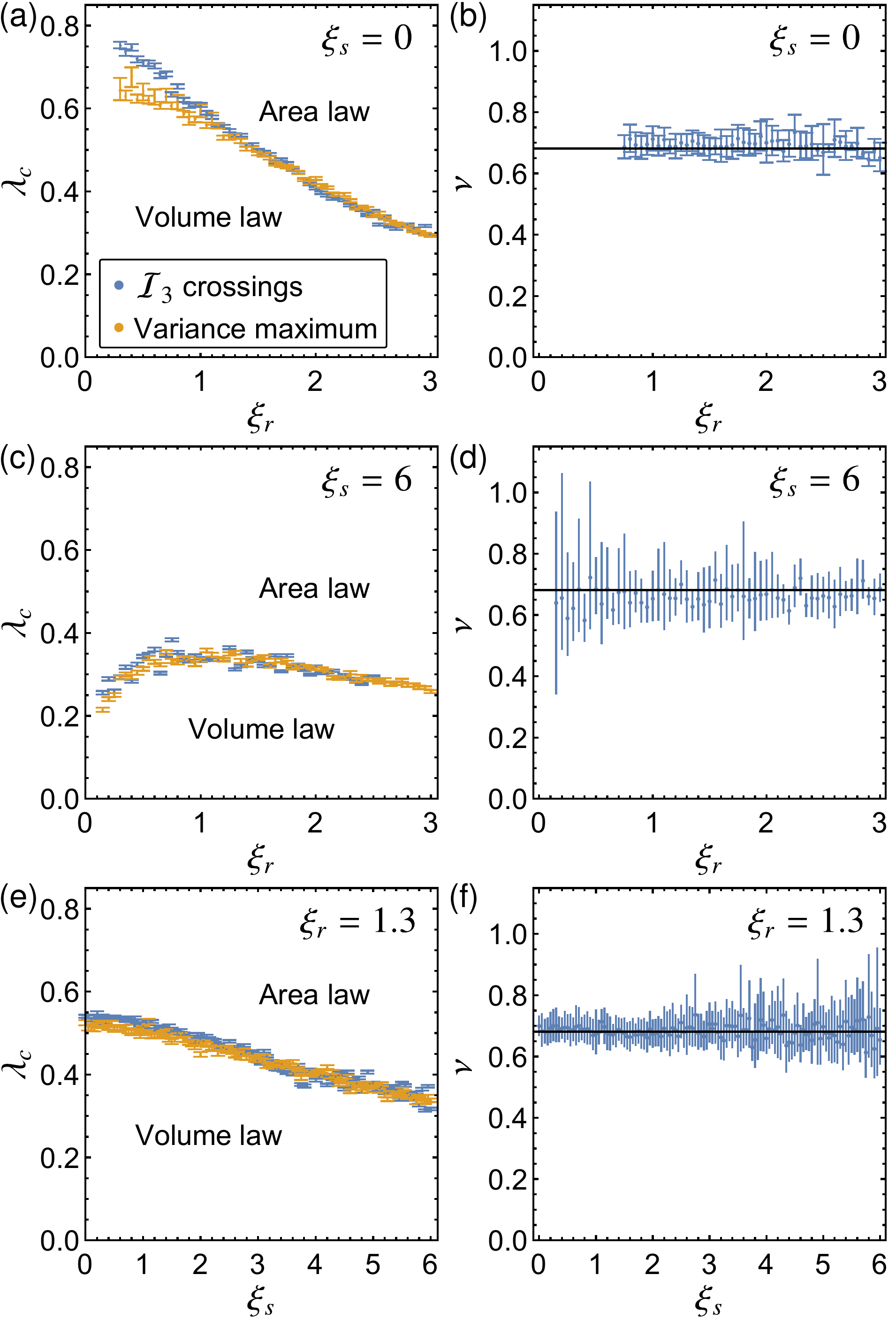}
  \caption{Universality of the critical exponent $\nu$ of the correlation length. Panels (a) and (c) show the critical measurement strength $\lambda_c$ as a function of the noise strength $\xi_r$ at vanishing disorder ($\xi_s=0$) and finite disorder ($\xi_s=6$) respectively extrapolated to the thermodynamic limit based on the location of maximal variance (orange) and TMI crossings (blue). Panel (e) shows the critical measurement strength as a function of disorder strength $\xi_s$ for $\xi_r=1.3$. Panels (b), (d) and (f) show the critical exponent $\nu$ for the three cases, extracted by collapsing the data according to the ansatz \eqref{eq:entropy_scaling_ansatz}. The black lines indicate the weighted average $0.685(5)$ of all extracted exponents.}
  \label{fig:critical_exponent}
\end{figure}

So far, we have focused on the close links between the measurement-induced steady-state entanglement transition and the transient unitary entanglement dynamics. To examine if this connection is overshadowed by hidden trends, we now turn to the finite-size analysis of the critical scaling around the transition point.

For this, we assume that the entanglement entropy adheres to the scaling ansatz developed in Refs.~\cite{Skinner2018, Li2019}, which removes the logarithmic divergence the entropy is expected to exhibit at the critical point:
\begin{equation}
  \bar{S}_\infty(L,\lambda) - \bar{S}_\infty(L,\lambda_c) = F\left[(\lambda - \lambda_c)L^{1/\nu}\right],
  \label{eq:entropy_scaling_ansatz}
\end{equation}
where $F$ is an unknown one-parameter scaling function.
Our goal is to extract the critical exponent of the correlation length, $\nu$, from a data collapse of the steady-state entanglement entropy $\bar{S}_\infty$.
We perform this analysis for varying $\xi_r$ and no static disorder ($\xi_s=0$), as well as for varying $\xi_s$ with $\xi_r = 1.3$.
To start out, we determine the extrapolated critical measurement strength  $\lambda_c$ in  the thermodynamic limit, following the two procedures outlined in Sec.~\ref{sec:Extracting_Lambda_Section} involving the maximal CRV and the crossing of the TMI.  As shown in Fig.~\ref{fig:critical_exponent}(a),
at $\xi_s=0$ the two methods are in excellent agreement when the noise $\xi_r \gtrsim 0.7$, where the statistical uncertainties of the extrapolated values are furthermore small. 
The extrapolation of the critical measurement strength in the thermodynamic limit for $\xi_s=6$ is reported in Fig~\ref{fig:critical_exponent}(c).  The results confirm that the non-monotonous dependence of $\lambda_c(\xi_r)$ and the shift of the maximum due to static noise persist in the thermodynamic limit. Notably, the excellent agreement  between the two extrapolation procedures applies to all values of $\xi_s$, as shown in panel (e).

To obtain the critical exponent, we take $\lambda_c$ from the extrapolation of maximal CRV, limiting it to the described range where it agrees well with the estimate from the TMI crossings.
Next, we carry out a polynomial fit of $\bar{S}_\infty(L, \lambda)$ for each system size $L$, and use this to extract an estimate of $\bar{S}_\infty(L, \lambda_c)$. Following this, we rescale the data as $y = \bar{S}_\infty(L, \lambda) - \bar{S}_\infty(L, \lambda_c)$ and $x = (\lambda - \lambda_c) L^{1 / \nu}$, and make a polynomial fit of order $m$ to the data near $\lambda = \lambda_c$. The quality of the collapse is measured by the reduced chi-squared $\chi^2(\nu)$ of the fit, averaged over $m=5,...,8$. Finally, minimizing $\chi^2(\nu)$ gives us an estimate of $\nu$, with the errors estimated as a band within $\chi^2(\nu) < 2 \chi^2(\nu_\text{min})$.

Figs.~\ref{fig:critical_exponent}(b), (d) and (f) show the outcome of the above procedure. Within error bars, the extracted exponent is constant over the considered range of $\xi_r$ and $\xi_s$, with a weighted average of $\nu = 0.685(5)$. This strongly suggests that systems with different $\xi_r$ and $\xi_s$ all belong to the same universality class. Moreover, the obtained value of $\nu$  is close to the exponents found for models with completely random stochastic unitary evolution protocols~\cite{Szyniszewski2020universality}, where $\nu \approx 0.70(1)$ for continuous measurements and $\nu \approx 0.80(3)$  for stroboscopic weak measurements.
Notably, these estimates differ significantly from the critical exponent found in a model with stroboscopic \emph{projective} measurements, where $\nu \approx 1.2(2)$ for Haar unitary evolution and $\nu \approx 1.24(7)$ for stabilizer circuits~\cite{Zabalo2020}.
This indicates that the entanglement transition induced by continuous measurements displays a large degree of universality, with all relevant features captured by the single parameter $\lambda_c$, and further supports our suggestion that this parameter provides reliable insights into the detailed entanglement dynamics of specific physical systems.

\section{Conclusions}
\label{sec:Conclusions}
In summary, we analyzed the entanglement dynamics of a one-dimensional many-body system evolving under
local continuous measurements and unitary dynamics combining interactions, temporal noise, and spatial background disorder.
The unitary dynamics is generated by a local Hamiltonian in the form of a spin chain, where fixed interactions entangle neighboring spins, while two independent parameters control the strength of temporal fluctuations and spatial disorder. The steady-state entanglement entropy displays a measurement-induced transition from extensive volume-law scaling to area-law scaling,  which we analyzed in detail based on finite-size scaling of the averaged steady-state entropy, its cross-realization variance, and the tripartite mutual information.

The obtained phase diagram  in the parameter space of measurement strength, time fluctuation and spatial disorder
reveals detailed features of the processes governing  the unitary entanglement dynamics.
As intuitively expected from many-body localization,
static disorder induces a trend towards states with limited entanglement, resulting in a systematic reduction of the critical measurement strength required to induce the transition to area-law entanglement.
However, we find that the critical measurement strength depends non-monotonically on the temporal fluctuations, supporting a picture where weak noise facilitates the spreading of entanglement via delocalization, while strong noise suppresses the efficiency by which interactions entangle the spins.  These steady-state features are well reflected by the transient unitary entanglement dynamics out of a separable state. The universal aspects of this connection are further  underpinned by a finite-size analysis of the critical scaling, where the obtained critical scaling exponent is independent of the noise and disorder strength, and takes a value compatible with the universality class of the entanglement transition in completely stochastically evolving continuously measured systems.

Our findings show that the measurement-induced entanglement transition reflects specific features of the physical processes participating in the unitary entanglement dynamics. This raises hopes to attain a microscopic understanding of this transition in specific physical settings, and opens up the possibility of using this transition as a diagnostic tool to infer details of transient entanglement dynamics via a steady-state phase transition.

\begin{acknowledgments}
This research was funded by EPSRC via Grant No.~EP/P010180/1. M.S.\ was also funded by the European Research Council (ERC) under the European Union's Horizon 2020 Research and Innovation programme (Grant agreement No.~853368). T.B. was also funded by the EPSRC via Grant No. EP/T518062/1. Computer time for this project was provided by the High-End Computing (HEC) facility at Lancaster University.
\end{acknowledgments}

\bibliography{Biblio}

\begin{thebibliography}{83}%
\makeatletter
\providecommand \@ifxundefined [1]{%
 \@ifx{#1\undefined}
}%
\providecommand \@ifnum [1]{%
 \ifnum #1\expandafter \@firstoftwo
 \else \expandafter \@secondoftwo
 \fi
}%
\providecommand \@ifx [1]{%
 \ifx #1\expandafter \@firstoftwo
 \else \expandafter \@secondoftwo
 \fi
}%
\providecommand \natexlab [1]{#1}%
\providecommand \enquote  [1]{``#1''}%
\providecommand \bibnamefont  [1]{#1}%
\providecommand \bibfnamefont [1]{#1}%
\providecommand \citenamefont [1]{#1}%
\providecommand \href@noop [0]{\@secondoftwo}%
\providecommand \href [0]{\begingroup \@sanitize@url \@href}%
\providecommand \@href[1]{\@@startlink{#1}\@@href}%
\providecommand \@@href[1]{\endgroup#1\@@endlink}%
\providecommand \@sanitize@url [0]{\catcode `\\12\catcode `\$12\catcode
  `\&12\catcode `\#12\catcode `\^12\catcode `\_12\catcode `\%12\relax}%
\providecommand \@@startlink[1]{}%
\providecommand \@@endlink[0]{}%
\providecommand \url  [0]{\begingroup\@sanitize@url \@url }%
\providecommand \@url [1]{\endgroup\@href {#1}{\urlprefix }}%
\providecommand \urlprefix  [0]{URL }%
\providecommand \Eprint [0]{\href }%
\providecommand \doibase [0]{https://doi.org/}%
\providecommand \selectlanguage [0]{\@gobble}%
\providecommand \bibinfo  [0]{\@secondoftwo}%
\providecommand \bibfield  [0]{\@secondoftwo}%
\providecommand \translation [1]{[#1]}%
\providecommand \BibitemOpen [0]{}%
\providecommand \bibitemStop [0]{}%
\providecommand \bibitemNoStop [0]{.\EOS\space}%
\providecommand \EOS [0]{\spacefactor3000\relax}%
\providecommand \BibitemShut  [1]{\csname bibitem#1\endcsname}%
\let\auto@bib@innerbib\@empty
\bibitem [{\citenamefont {Deutsch}(1991)}]{Deutsch1991}%
  \BibitemOpen
  \bibfield  {author} {\bibinfo {author} {\bibfnamefont {J.~M.}\ \bibnamefont
  {Deutsch}},\ }\bibfield  {title} {\bibinfo {title} {Quantum statistical
  mechanics in a closed system},\ }\href
  {https://doi.org/10.1103/PhysRevA.43.2046} {\bibfield  {journal} {\bibinfo
  {journal} {Phys. Rev. A}\ }\textbf {\bibinfo {volume} {43}},\ \bibinfo
  {pages} {2046} (\bibinfo {year} {1991})}\BibitemShut {NoStop}%
\bibitem [{\citenamefont {Srednicki}(1994)}]{Srednicki1994}%
  \BibitemOpen
  \bibfield  {author} {\bibinfo {author} {\bibfnamefont {M.}~\bibnamefont
  {Srednicki}},\ }\bibfield  {title} {\bibinfo {title} {Chaos and quantum
  thermalization},\ }\href {https://doi.org/10.1103/PhysRevE.50.888} {\bibfield
   {journal} {\bibinfo  {journal} {Phys. Rev. E}\ }\textbf {\bibinfo {volume}
  {50}},\ \bibinfo {pages} {888} (\bibinfo {year} {1994})}\BibitemShut
  {NoStop}%
\bibitem [{\citenamefont {D'Alessio}\ \emph {et~al.}(2016)\citenamefont
  {D'Alessio}, \citenamefont {Kafri}, \citenamefont {Polkovnikov},\ and\
  \citenamefont {Rigol}}]{DAlessio2016}%
  \BibitemOpen
  \bibfield  {author} {\bibinfo {author} {\bibfnamefont {L.}~\bibnamefont
  {D'Alessio}}, \bibinfo {author} {\bibfnamefont {Y.}~\bibnamefont {Kafri}},
  \bibinfo {author} {\bibfnamefont {A.}~\bibnamefont {Polkovnikov}},\ and\
  \bibinfo {author} {\bibfnamefont {M.}~\bibnamefont {Rigol}},\ }\bibfield
  {title} {\bibinfo {title} {From quantum chaos and eigenstate thermalization
  to statistical mechanics and thermodynamics},\ }\href
  {https://doi.org/10.1080/00018732.2016.1198134} {\bibfield  {journal}
  {\bibinfo  {journal} {Adv. Phys.}\ }\textbf {\bibinfo {volume} {65}},\
  \bibinfo {pages} {239} (\bibinfo {year} {2016})}\BibitemShut {NoStop}%
\bibitem [{\citenamefont {Borgonovi}\ \emph {et~al.}(2016)\citenamefont
  {Borgonovi}, \citenamefont {Izrailev}, \citenamefont {Santos},\ and\
  \citenamefont {Zelevinsky}}]{Borgonovi2016}%
  \BibitemOpen
  \bibfield  {author} {\bibinfo {author} {\bibfnamefont {F.}~\bibnamefont
  {Borgonovi}}, \bibinfo {author} {\bibfnamefont {F.~M.}\ \bibnamefont
  {Izrailev}}, \bibinfo {author} {\bibfnamefont {L.~F.}\ \bibnamefont
  {Santos}},\ and\ \bibinfo {author} {\bibfnamefont {V.~G.}\ \bibnamefont
  {Zelevinsky}},\ }\bibfield  {title} {\bibinfo {title} {Quantum chaos and
  thermalization in isolated systems of interacting particles},\ }\href
  {https://doi.org/10.1016/j.physrep.2016.02.005} {\bibfield  {journal}
  {\bibinfo  {journal} {Phys. Rep.}\ }\textbf {\bibinfo {volume} {626}},\
  \bibinfo {pages} {1} (\bibinfo {year} {2016})}\BibitemShut {NoStop}%
\bibitem [{\citenamefont {Garrison}\ and\ \citenamefont
  {Grover}(2018)}]{garrison2018does}%
  \BibitemOpen
  \bibfield  {author} {\bibinfo {author} {\bibfnamefont {J.~R.}\ \bibnamefont
  {Garrison}}\ and\ \bibinfo {author} {\bibfnamefont {T.}~\bibnamefont
  {Grover}},\ }\bibfield  {title} {\bibinfo {title} {Does a single eigenstate
  encode the full {Hamiltonian}?},\ }\href
  {https://doi.org/10.1103/PhysRevX.8.021026} {\bibfield  {journal} {\bibinfo
  {journal} {Phys. Rev. X}\ }\textbf {\bibinfo {volume} {8}},\ \bibinfo {pages}
  {021026} (\bibinfo {year} {2018})}\BibitemShut {NoStop}%
\bibitem [{\citenamefont {Kim}\ \emph {et~al.}(2014)\citenamefont {Kim},
  \citenamefont {Ikeda},\ and\ \citenamefont {Huse}}]{kim2014testing}%
  \BibitemOpen
  \bibfield  {author} {\bibinfo {author} {\bibfnamefont {H.}~\bibnamefont
  {Kim}}, \bibinfo {author} {\bibfnamefont {T.~N.}\ \bibnamefont {Ikeda}},\
  and\ \bibinfo {author} {\bibfnamefont {D.~A.}\ \bibnamefont {Huse}},\
  }\bibfield  {title} {\bibinfo {title} {Testing whether all eigenstates obey
  the eigenstate thermalization hypothesis},\ }\href
  {https://doi.org/10.1103/PhysRevE.90.052105} {\bibfield  {journal} {\bibinfo
  {journal} {Phys. Rev. E}\ }\textbf {\bibinfo {volume} {90}},\ \bibinfo
  {pages} {052105} (\bibinfo {year} {2014})}\BibitemShut {NoStop}%
\bibitem [{\citenamefont {Bao}\ and\ \citenamefont
  {Cheng}(2019)}]{Bao2019Eigenstate}%
  \BibitemOpen
  \bibfield  {author} {\bibinfo {author} {\bibfnamefont {N.}~\bibnamefont
  {Bao}}\ and\ \bibinfo {author} {\bibfnamefont {N.}~\bibnamefont {Cheng}},\
  }\bibfield  {title} {\bibinfo {title} {Eigenstate thermalization hypothesis
  and approximate quantum error correction},\ }\href
  {https://doi.org/10.1007/JHEP08(2019)152} {\bibfield  {journal} {\bibinfo
  {journal} {J. High Energy Phys.}\ }\textbf {\bibinfo {volume} {2019}}\bibinfo
   {number} { (8)},\ \bibinfo {pages} {152}}\BibitemShut {NoStop}%
\bibitem [{\citenamefont {Nandkishore}\ and\ \citenamefont
  {Huse}(2015)}]{nandkishore2014review}%
  \BibitemOpen
\bibfield  {number} {  }\bibfield  {author} {\bibinfo {author} {\bibfnamefont
  {R.}~\bibnamefont {Nandkishore}}\ and\ \bibinfo {author} {\bibfnamefont
  {D.~A.}\ \bibnamefont {Huse}},\ }\bibfield  {title} {\bibinfo {title}
  {Many-body localization and thermalization in quantum statistical
  mechanics},\ }\href
  {https://doi.org/10.1146/annurev-conmatphys-031214-014726} {\bibfield
  {journal} {\bibinfo  {journal} {Annu. Rev. Condens. Matter Phys.}\ }\textbf
  {\bibinfo {volume} {6}},\ \bibinfo {pages} {15} (\bibinfo {year}
  {2015})}\BibitemShut {NoStop}%
\bibitem [{\citenamefont {Abanin}\ and\ \citenamefont
  {Papi{\'c}}(2017)}]{abanin2017recent}%
  \BibitemOpen
  \bibfield  {author} {\bibinfo {author} {\bibfnamefont {D.~A.}\ \bibnamefont
  {Abanin}}\ and\ \bibinfo {author} {\bibfnamefont {Z.}~\bibnamefont
  {Papi{\'c}}},\ }\bibfield  {title} {\bibinfo {title} {Recent progress in
  many-body localization},\ }\href {https://doi.org/10.1002/andp.201700169}
  {\bibfield  {journal} {\bibinfo  {journal} {Ann. Phys. (Berl.)}\ }\textbf
  {\bibinfo {volume} {529}},\ \bibinfo {pages} {1700169} (\bibinfo {year}
  {2017})}\BibitemShut {NoStop}%
\bibitem [{\citenamefont {Abanin}\ \emph {et~al.}(2019)\citenamefont {Abanin},
  \citenamefont {Altman}, \citenamefont {Bloch},\ and\ \citenamefont
  {Serbyn}}]{Abanin2019Review}%
  \BibitemOpen
  \bibfield  {author} {\bibinfo {author} {\bibfnamefont {D.~A.}\ \bibnamefont
  {Abanin}}, \bibinfo {author} {\bibfnamefont {E.}~\bibnamefont {Altman}},
  \bibinfo {author} {\bibfnamefont {I.}~\bibnamefont {Bloch}},\ and\ \bibinfo
  {author} {\bibfnamefont {M.}~\bibnamefont {Serbyn}},\ }\bibfield  {title}
  {\bibinfo {title} {Colloquium: Many-body localization, thermalization, and
  entanglement},\ }\href {https://doi.org/10.1103/RevModPhys.91.021001}
  {\bibfield  {journal} {\bibinfo  {journal} {Rev. Mod. Phys.}\ }\textbf
  {\bibinfo {volume} {91}},\ \bibinfo {pages} {021001} (\bibinfo {year}
  {2019})}\BibitemShut {NoStop}%
\bibitem [{\citenamefont {Pal}\ and\ \citenamefont {Huse}(2010)}]{pal2010many}%
  \BibitemOpen
  \bibfield  {author} {\bibinfo {author} {\bibfnamefont {A.}~\bibnamefont
  {Pal}}\ and\ \bibinfo {author} {\bibfnamefont {D.~A.}\ \bibnamefont {Huse}},\
  }\bibfield  {title} {\bibinfo {title} {Many-body localization phase
  transition},\ }\href {https://doi.org/10.1103/PhysRevB.82.174411} {\bibfield
  {journal} {\bibinfo  {journal} {Phys. Rev. B}\ }\textbf {\bibinfo {volume}
  {82}},\ \bibinfo {pages} {174411} (\bibinfo {year} {2010})}\BibitemShut
  {NoStop}%
\bibitem [{\citenamefont {\ifmmode \check{Z}\else
  \v{Z}\fi{}nidari\ifmmode~\check{c}\else \v{c}\fi{}}\ \emph
  {et~al.}(2008)\citenamefont {\ifmmode \check{Z}\else
  \v{Z}\fi{}nidari\ifmmode~\check{c}\else \v{c}\fi{}}, \citenamefont {Prosen},\
  and\ \citenamefont {Prelov\ifmmode~\check{s}\else
  \v{s}\fi{}ek}}]{Prosen2008Many}%
  \BibitemOpen
  \bibfield  {author} {\bibinfo {author} {\bibfnamefont {M.}~\bibnamefont
  {\ifmmode \check{Z}\else \v{Z}\fi{}nidari\ifmmode~\check{c}\else
  \v{c}\fi{}}}, \bibinfo {author} {\bibfnamefont {T.}~\bibnamefont {Prosen}},\
  and\ \bibinfo {author} {\bibfnamefont {P.}~\bibnamefont
  {Prelov\ifmmode~\check{s}\else \v{s}\fi{}ek}},\ }\bibfield  {title} {\bibinfo
  {title} {Many-body localization in the {Heisenberg} {$XXZ$} magnet in a
  random field},\ }\href {https://doi.org/10.1103/PhysRevB.77.064426}
  {\bibfield  {journal} {\bibinfo  {journal} {Phys. Rev. B}\ }\textbf {\bibinfo
  {volume} {77}},\ \bibinfo {pages} {064426} (\bibinfo {year}
  {2008})}\BibitemShut {NoStop}%
\bibitem [{\citenamefont {Basko}\ \emph {et~al.}(2006)\citenamefont {Basko},
  \citenamefont {Aleiner},\ and\ \citenamefont {Altshuler}}]{Basko2006}%
  \BibitemOpen
  \bibfield  {author} {\bibinfo {author} {\bibfnamefont {D.~M.}\ \bibnamefont
  {Basko}}, \bibinfo {author} {\bibfnamefont {I.~L.}\ \bibnamefont {Aleiner}},\
  and\ \bibinfo {author} {\bibfnamefont {B.~L.}\ \bibnamefont {Altshuler}},\
  }\bibfield  {title} {\bibinfo {title} {Metal--insulator transition in a
  weakly interacting many-electron system with localized single-particle
  states},\ }\href {https://doi.org/10.1016/j.aop.2005.11.014} {\bibfield
  {journal} {\bibinfo  {journal} {Ann. Phys.}\ }\textbf {\bibinfo {volume}
  {321}},\ \bibinfo {pages} {1126} (\bibinfo {year} {2006})}\BibitemShut
  {NoStop}%
\bibitem [{\citenamefont {Gornyi}\ \emph {et~al.}(2005)\citenamefont {Gornyi},
  \citenamefont {Mirlin},\ and\ \citenamefont {Polyakov}}]{Gornyi2005}%
  \BibitemOpen
  \bibfield  {author} {\bibinfo {author} {\bibfnamefont {I.~V.}\ \bibnamefont
  {Gornyi}}, \bibinfo {author} {\bibfnamefont {A.~D.}\ \bibnamefont {Mirlin}},\
  and\ \bibinfo {author} {\bibfnamefont {D.~G.}\ \bibnamefont {Polyakov}},\
  }\bibfield  {title} {\bibinfo {title} {Interacting electrons in disordered
  wires: Anderson localization and low-{$T$} transport},\ }\href
  {https://doi.org/10.1103/PhysRevLett.95.206603} {\bibfield  {journal}
  {\bibinfo  {journal} {Phys. Rev. Lett.}\ }\textbf {\bibinfo {volume} {95}},\
  \bibinfo {pages} {206603} (\bibinfo {year} {2005})}\BibitemShut {NoStop}%
\bibitem [{\citenamefont {Altman}\ and\ \citenamefont
  {Vosk}(2015)}]{Altman2015}%
  \BibitemOpen
  \bibfield  {author} {\bibinfo {author} {\bibfnamefont {E.}~\bibnamefont
  {Altman}}\ and\ \bibinfo {author} {\bibfnamefont {R.}~\bibnamefont {Vosk}},\
  }\bibfield  {title} {\bibinfo {title} {Universal dynamics and renormalization
  in many-body-localized systems},\ }\href
  {https://doi.org/10.1146/annurev-conmatphys-031214-014701} {\bibfield
  {journal} {\bibinfo  {journal} {Annu. Rev. Condens. Matter Phys.}\ }\textbf
  {\bibinfo {volume} {6}},\ \bibinfo {pages} {383} (\bibinfo {year}
  {2015})}\BibitemShut {NoStop}%
\bibitem [{\citenamefont {Chandran}\ \emph {et~al.}(2015)\citenamefont
  {Chandran}, \citenamefont {Kim}, \citenamefont {Vidal},\ and\ \citenamefont
  {Abanin}}]{chandran2015constructing}%
  \BibitemOpen
  \bibfield  {author} {\bibinfo {author} {\bibfnamefont {A.}~\bibnamefont
  {Chandran}}, \bibinfo {author} {\bibfnamefont {I.~H.}\ \bibnamefont {Kim}},
  \bibinfo {author} {\bibfnamefont {G.}~\bibnamefont {Vidal}},\ and\ \bibinfo
  {author} {\bibfnamefont {D.~A.}\ \bibnamefont {Abanin}},\ }\bibfield  {title}
  {\bibinfo {title} {Constructing local integrals of motion in the many-body
  localized phase},\ }\href {https://doi.org/10.1103/PhysRevB.91.085425}
  {\bibfield  {journal} {\bibinfo  {journal} {Phys. Rev. B}\ }\textbf {\bibinfo
  {volume} {91}},\ \bibinfo {pages} {085425} (\bibinfo {year}
  {2015})}\BibitemShut {NoStop}%
\bibitem [{\citenamefont {Luitz}\ \emph {et~al.}(2015)\citenamefont {Luitz},
  \citenamefont {Laflorencie},\ and\ \citenamefont {Alet}}]{luitz2015many}%
  \BibitemOpen
  \bibfield  {author} {\bibinfo {author} {\bibfnamefont {D.~J.}\ \bibnamefont
  {Luitz}}, \bibinfo {author} {\bibfnamefont {N.}~\bibnamefont {Laflorencie}},\
  and\ \bibinfo {author} {\bibfnamefont {F.}~\bibnamefont {Alet}},\ }\bibfield
  {title} {\bibinfo {title} {Many-body localization edge in the random-field
  {Heisenberg} chain},\ }\href {https://doi.org/10.1103/PhysRevB.91.081103}
  {\bibfield  {journal} {\bibinfo  {journal} {Phys. Rev. B}\ }\textbf {\bibinfo
  {volume} {91}},\ \bibinfo {pages} {081103} (\bibinfo {year}
  {2015})}\BibitemShut {NoStop}%
\bibitem [{\citenamefont {Parameswaran}\ and\ \citenamefont
  {Vasseur}(2018)}]{Parameswaran2018}%
  \BibitemOpen
  \bibfield  {author} {\bibinfo {author} {\bibfnamefont {S.~A.}\ \bibnamefont
  {Parameswaran}}\ and\ \bibinfo {author} {\bibfnamefont {R.}~\bibnamefont
  {Vasseur}},\ }\bibfield  {title} {\bibinfo {title} {Many-body localization,
  symmetry and topology},\ }\href {https://doi.org/10.1088/1361-6633/aac9ed}
  {\bibfield  {journal} {\bibinfo  {journal} {Rep. Prog. Phys.}\ }\textbf
  {\bibinfo {volume} {81}},\ \bibinfo {pages} {082501} (\bibinfo {year}
  {2018})}\BibitemShut {NoStop}%
\bibitem [{\citenamefont {Li}\ \emph {et~al.}(2018)\citenamefont {Li},
  \citenamefont {Chen},\ and\ \citenamefont {Fisher}}]{Li2018}%
  \BibitemOpen
  \bibfield  {author} {\bibinfo {author} {\bibfnamefont {Y.}~\bibnamefont
  {Li}}, \bibinfo {author} {\bibfnamefont {X.}~\bibnamefont {Chen}},\ and\
  \bibinfo {author} {\bibfnamefont {M.~P.~A.}\ \bibnamefont {Fisher}},\
  }\bibfield  {title} {\bibinfo {title} {{Quantum Zeno effect and the many-body
  entanglement transition}},\ }\href
  {https://doi.org/10.1103/PhysRevB.98.205136} {\bibfield  {journal} {\bibinfo
  {journal} {Phys. Rev. B}\ }\textbf {\bibinfo {volume} {98}},\ \bibinfo
  {pages} {205136} (\bibinfo {year} {2018})}\BibitemShut {NoStop}%
\bibitem [{\citenamefont {Chan}\ \emph {et~al.}(2019)\citenamefont {Chan},
  \citenamefont {Nandkishore}, \citenamefont {Pretko},\ and\ \citenamefont
  {Smith}}]{Chan2018}%
  \BibitemOpen
  \bibfield  {author} {\bibinfo {author} {\bibfnamefont {A.}~\bibnamefont
  {Chan}}, \bibinfo {author} {\bibfnamefont {R.~M.}\ \bibnamefont
  {Nandkishore}}, \bibinfo {author} {\bibfnamefont {M.}~\bibnamefont
  {Pretko}},\ and\ \bibinfo {author} {\bibfnamefont {G.}~\bibnamefont
  {Smith}},\ }\bibfield  {title} {\bibinfo {title} {Unitary-projective
  entanglement dynamics},\ }\href {https://doi.org/10.1103/PhysRevB.99.224307}
  {\bibfield  {journal} {\bibinfo  {journal} {Phys. Rev. B}\ }\textbf {\bibinfo
  {volume} {99}},\ \bibinfo {pages} {224307} (\bibinfo {year}
  {2019})}\BibitemShut {NoStop}%
\bibitem [{\citenamefont {Skinner}\ \emph {et~al.}(2019)\citenamefont
  {Skinner}, \citenamefont {Ruhman},\ and\ \citenamefont
  {Nahum}}]{Skinner2018}%
  \BibitemOpen
  \bibfield  {author} {\bibinfo {author} {\bibfnamefont {B.}~\bibnamefont
  {Skinner}}, \bibinfo {author} {\bibfnamefont {J.}~\bibnamefont {Ruhman}},\
  and\ \bibinfo {author} {\bibfnamefont {A.}~\bibnamefont {Nahum}},\ }\bibfield
   {title} {\bibinfo {title} {Measurement-induced phase transitions in the
  dynamics of entanglement},\ }\href
  {https://doi.org/10.1103/PhysRevX.9.031009} {\bibfield  {journal} {\bibinfo
  {journal} {Phys. Rev. X}\ }\textbf {\bibinfo {volume} {9}},\ \bibinfo {pages}
  {031009} (\bibinfo {year} {2019})}\BibitemShut {NoStop}%
\bibitem [{\citenamefont {Li}\ \emph {et~al.}(2019)\citenamefont {Li},
  \citenamefont {Chen},\ and\ \citenamefont {Fisher}}]{Li2019}%
  \BibitemOpen
  \bibfield  {author} {\bibinfo {author} {\bibfnamefont {Y.}~\bibnamefont
  {Li}}, \bibinfo {author} {\bibfnamefont {X.}~\bibnamefont {Chen}},\ and\
  \bibinfo {author} {\bibfnamefont {M.~P.~A.}\ \bibnamefont {Fisher}},\
  }\bibfield  {title} {\bibinfo {title} {Measurement-driven entanglement
  transition in hybrid quantum circuits},\ }\href
  {https://doi.org/10.1103/PhysRevB.100.134306} {\bibfield  {journal} {\bibinfo
   {journal} {Phys. Rev. B}\ }\textbf {\bibinfo {volume} {100}},\ \bibinfo
  {pages} {134306} (\bibinfo {year} {2019})}\BibitemShut {NoStop}%
\bibitem [{\citenamefont {Szyniszewski}\ \emph {et~al.}(2019)\citenamefont
  {Szyniszewski}, \citenamefont {Romito},\ and\ \citenamefont
  {Schomerus}}]{Szyniszewski2019}%
  \BibitemOpen
  \bibfield  {author} {\bibinfo {author} {\bibfnamefont {M.}~\bibnamefont
  {Szyniszewski}}, \bibinfo {author} {\bibfnamefont {A.}~\bibnamefont
  {Romito}},\ and\ \bibinfo {author} {\bibfnamefont {H.}~\bibnamefont
  {Schomerus}},\ }\bibfield  {title} {\bibinfo {title} {Entanglement transition
  from variable-strength weak measurements},\ }\href
  {https://doi.org/10.1103/PhysRevB.100.064204} {\bibfield  {journal} {\bibinfo
   {journal} {Phys. Rev. B}\ }\textbf {\bibinfo {volume} {100}},\ \bibinfo
  {pages} {064204} (\bibinfo {year} {2019})}\BibitemShut {NoStop}%
\bibitem [{\citenamefont {Zabalo}\ \emph {et~al.}(2020)\citenamefont {Zabalo},
  \citenamefont {Gullans}, \citenamefont {Wilson}, \citenamefont
  {Gopalakrishnan}, \citenamefont {Huse},\ and\ \citenamefont
  {Pixley}}]{Zabalo2020}%
  \BibitemOpen
  \bibfield  {author} {\bibinfo {author} {\bibfnamefont {A.}~\bibnamefont
  {Zabalo}}, \bibinfo {author} {\bibfnamefont {M.~J.}\ \bibnamefont {Gullans}},
  \bibinfo {author} {\bibfnamefont {J.~H.}\ \bibnamefont {Wilson}}, \bibinfo
  {author} {\bibfnamefont {S.}~\bibnamefont {Gopalakrishnan}}, \bibinfo
  {author} {\bibfnamefont {D.~A.}\ \bibnamefont {Huse}},\ and\ \bibinfo
  {author} {\bibfnamefont {J.~H.}\ \bibnamefont {Pixley}},\ }\bibfield  {title}
  {\bibinfo {title} {Critical properties of the measurement-induced transition
  in random quantum circuits},\ }\href
  {https://doi.org/10.1103/PhysRevB.101.060301} {\bibfield  {journal} {\bibinfo
   {journal} {Phys. Rev. B}\ }\textbf {\bibinfo {volume} {101}},\ \bibinfo
  {pages} {060301(R)} (\bibinfo {year} {2020})}\BibitemShut {NoStop}%
\bibitem [{\citenamefont {Napp}\ \emph {et~al.}(2019)\citenamefont {Napp},
  \citenamefont {Placa}, \citenamefont {Dalzell}, \citenamefont {Brandao},\
  and\ \citenamefont {Harrow}}]{Napp2019}%
  \BibitemOpen
  \bibfield  {author} {\bibinfo {author} {\bibfnamefont {J.}~\bibnamefont
  {Napp}}, \bibinfo {author} {\bibfnamefont {R.~L.~L.}\ \bibnamefont {Placa}},
  \bibinfo {author} {\bibfnamefont {A.~M.}\ \bibnamefont {Dalzell}}, \bibinfo
  {author} {\bibfnamefont {F.~G. S.~L.}\ \bibnamefont {Brandao}},\ and\
  \bibinfo {author} {\bibfnamefont {A.~W.}\ \bibnamefont {Harrow}},\ }\bibfield
   {title} {\bibinfo {title} {Efficient classical simulation of random shallow
  {2D} quantum circuits},\ }\Eprint {https://arxiv.org/abs/2001.00021}
  {arXiv:2001.00021}  (\bibinfo {year} {2019})\BibitemShut {NoStop}%
\bibitem [{\citenamefont {Fan}\ \emph {et~al.}(2021)\citenamefont {Fan},
  \citenamefont {Vijay}, \citenamefont {Vishwanath},\ and\ \citenamefont
  {You}}]{Fan2021}%
  \BibitemOpen
  \bibfield  {author} {\bibinfo {author} {\bibfnamefont {R.}~\bibnamefont
  {Fan}}, \bibinfo {author} {\bibfnamefont {S.}~\bibnamefont {Vijay}}, \bibinfo
  {author} {\bibfnamefont {A.}~\bibnamefont {Vishwanath}},\ and\ \bibinfo
  {author} {\bibfnamefont {Y.-Z.}\ \bibnamefont {You}},\ }\bibfield  {title}
  {\bibinfo {title} {Self-organized error correction in random unitary circuits
  with measurement},\ }\href {https://doi.org/10.1103/PhysRevB.103.174309}
  {\bibfield  {journal} {\bibinfo  {journal} {Phys. Rev. B}\ }\textbf {\bibinfo
  {volume} {103}},\ \bibinfo {pages} {174309} (\bibinfo {year}
  {2021})}\BibitemShut {NoStop}%
\bibitem [{\citenamefont {Gullans}\ and\ \citenamefont
  {Huse}(2020{\natexlab{a}})}]{Gullans2019purification}%
  \BibitemOpen
  \bibfield  {author} {\bibinfo {author} {\bibfnamefont {M.~J.}\ \bibnamefont
  {Gullans}}\ and\ \bibinfo {author} {\bibfnamefont {D.~A.}\ \bibnamefont
  {Huse}},\ }\bibfield  {title} {\bibinfo {title} {Dynamical purification phase
  transition induced by quantum measurements},\ }\href
  {https://doi.org/10.1103/PhysRevX.10.041020} {\bibfield  {journal} {\bibinfo
  {journal} {Phys. Rev. X}\ }\textbf {\bibinfo {volume} {10}},\ \bibinfo
  {pages} {041020} (\bibinfo {year} {2020}{\natexlab{a}})}\BibitemShut
  {NoStop}%
\bibitem [{\citenamefont {Bao}\ \emph {et~al.}(2020)\citenamefont {Bao},
  \citenamefont {Choi},\ and\ \citenamefont {Altman}}]{Bao2020}%
  \BibitemOpen
  \bibfield  {author} {\bibinfo {author} {\bibfnamefont {Y.}~\bibnamefont
  {Bao}}, \bibinfo {author} {\bibfnamefont {S.}~\bibnamefont {Choi}},\ and\
  \bibinfo {author} {\bibfnamefont {E.}~\bibnamefont {Altman}},\ }\bibfield
  {title} {\bibinfo {title} {Theory of the phase transition in random unitary
  circuits with measurements},\ }\href
  {https://doi.org/10.1103/PhysRevB.101.104301} {\bibfield  {journal} {\bibinfo
   {journal} {Phys. Rev. B}\ }\textbf {\bibinfo {volume} {101}},\ \bibinfo
  {pages} {104301} (\bibinfo {year} {2020})}\BibitemShut {NoStop}%
\bibitem [{\citenamefont {Bera}\ and\ \citenamefont
  {Singha~Roy}(2020)}]{Bera2020}%
  \BibitemOpen
  \bibfield  {author} {\bibinfo {author} {\bibfnamefont {A.}~\bibnamefont
  {Bera}}\ and\ \bibinfo {author} {\bibfnamefont {S.}~\bibnamefont
  {Singha~Roy}},\ }\bibfield  {title} {\bibinfo {title} {Growth of genuine
  multipartite entanglement in random unitary circuits},\ }\href
  {https://doi.org/10.1103/PhysRevA.102.062431} {\bibfield  {journal} {\bibinfo
   {journal} {Phys. Rev. A}\ }\textbf {\bibinfo {volume} {102}},\ \bibinfo
  {pages} {062431} (\bibinfo {year} {2020})}\BibitemShut {NoStop}%
\bibitem [{\citenamefont {Jian}\ \emph {et~al.}(2020)\citenamefont {Jian},
  \citenamefont {You}, \citenamefont {Vasseur},\ and\ \citenamefont
  {Ludwig}}]{Jian2020}%
  \BibitemOpen
  \bibfield  {author} {\bibinfo {author} {\bibfnamefont {C.-M.}\ \bibnamefont
  {Jian}}, \bibinfo {author} {\bibfnamefont {Y.-Z.}\ \bibnamefont {You}},
  \bibinfo {author} {\bibfnamefont {R.}~\bibnamefont {Vasseur}},\ and\ \bibinfo
  {author} {\bibfnamefont {A.~W.~W.}\ \bibnamefont {Ludwig}},\ }\bibfield
  {title} {\bibinfo {title} {Measurement-induced criticality in random quantum
  circuits},\ }\href {https://doi.org/10.1103/PhysRevB.101.104302} {\bibfield
  {journal} {\bibinfo  {journal} {Phys. Rev. B}\ }\textbf {\bibinfo {volume}
  {101}},\ \bibinfo {pages} {104302} (\bibinfo {year} {2020})}\BibitemShut
  {NoStop}%
\bibitem [{\citenamefont {Li}\ \emph {et~al.}(2020)\citenamefont {Li},
  \citenamefont {Chen}, \citenamefont {Ludwig},\ and\ \citenamefont
  {Fisher}}]{Li2020Conformal}%
  \BibitemOpen
  \bibfield  {author} {\bibinfo {author} {\bibfnamefont {Y.}~\bibnamefont
  {Li}}, \bibinfo {author} {\bibfnamefont {X.}~\bibnamefont {Chen}}, \bibinfo
  {author} {\bibfnamefont {A.~W.~W.}\ \bibnamefont {Ludwig}},\ and\ \bibinfo
  {author} {\bibfnamefont {M.~P.~A.}\ \bibnamefont {Fisher}},\ }\bibfield
  {title} {\bibinfo {title} {Conformal invariance and quantum non-locality in
  hybrid quantum circuits},\ }\Eprint {https://arxiv.org/abs/2003.12721}
  {arXiv:2003.12721}  (\bibinfo {year} {2020})\BibitemShut {NoStop}%
\bibitem [{\citenamefont {Szyniszewski}\ \emph {et~al.}(2020)\citenamefont
  {Szyniszewski}, \citenamefont {Romito},\ and\ \citenamefont
  {Schomerus}}]{Szyniszewski2020universality}%
  \BibitemOpen
  \bibfield  {author} {\bibinfo {author} {\bibfnamefont {M.}~\bibnamefont
  {Szyniszewski}}, \bibinfo {author} {\bibfnamefont {A.}~\bibnamefont
  {Romito}},\ and\ \bibinfo {author} {\bibfnamefont {H.}~\bibnamefont
  {Schomerus}},\ }\bibfield  {title} {\bibinfo {title} {Universality of
  entanglement transitions from stroboscopic to continuous measurements},\
  }\href {https://doi.org/10.1103/PhysRevLett.125.210602} {\bibfield  {journal}
  {\bibinfo  {journal} {Phys. Rev. Lett.}\ }\textbf {\bibinfo {volume} {125}},\
  \bibinfo {pages} {210602} (\bibinfo {year} {2020})}\BibitemShut {NoStop}%
\bibitem [{\citenamefont {Lopez-Piqueres}\ \emph {et~al.}(2020)\citenamefont
  {Lopez-Piqueres}, \citenamefont {Ware},\ and\ \citenamefont
  {Vasseur}}]{LopezPiqueres2020}%
  \BibitemOpen
  \bibfield  {author} {\bibinfo {author} {\bibfnamefont {J.}~\bibnamefont
  {Lopez-Piqueres}}, \bibinfo {author} {\bibfnamefont {B.}~\bibnamefont
  {Ware}},\ and\ \bibinfo {author} {\bibfnamefont {R.}~\bibnamefont
  {Vasseur}},\ }\bibfield  {title} {\bibinfo {title} {Mean-field entanglement
  transitions in random tree tensor networks},\ }\href
  {https://doi.org/10.1103/PhysRevB.102.064202} {\bibfield  {journal} {\bibinfo
   {journal} {Phys. Rev. B}\ }\textbf {\bibinfo {volume} {102}},\ \bibinfo
  {pages} {064202} (\bibinfo {year} {2020})}\BibitemShut {NoStop}%
\bibitem [{\citenamefont {Shtanko}\ \emph {et~al.}(2020)\citenamefont
  {Shtanko}, \citenamefont {Kharkov}, \citenamefont {Garc\'ia-Pintos},\ and\
  \citenamefont {Gorshkov}}]{Shtanko2020}%
  \BibitemOpen
  \bibfield  {author} {\bibinfo {author} {\bibfnamefont {O.}~\bibnamefont
  {Shtanko}}, \bibinfo {author} {\bibfnamefont {Y.~A.}\ \bibnamefont
  {Kharkov}}, \bibinfo {author} {\bibfnamefont {L.~P.}\ \bibnamefont
  {Garc\'ia-Pintos}},\ and\ \bibinfo {author} {\bibfnamefont {A.~V.}\
  \bibnamefont {Gorshkov}},\ }\bibfield  {title} {\bibinfo {title} {Classical
  models of entanglement in monitored random circuits},\ }\Eprint
  {https://arxiv.org/abs/2004.06736} {arXiv:2004.06736}  (\bibinfo {year}
  {2020})\BibitemShut {NoStop}%
\bibitem [{\citenamefont {Lavasani}\ \emph {et~al.}(2021)\citenamefont
  {Lavasani}, \citenamefont {Alavirad},\ and\ \citenamefont
  {Barkeshli}}]{Lavasani2021}%
  \BibitemOpen
  \bibfield  {author} {\bibinfo {author} {\bibfnamefont {A.}~\bibnamefont
  {Lavasani}}, \bibinfo {author} {\bibfnamefont {Y.}~\bibnamefont {Alavirad}},\
  and\ \bibinfo {author} {\bibfnamefont {M.}~\bibnamefont {Barkeshli}},\
  }\bibfield  {title} {\bibinfo {title} {Measurement-induced topological
  entanglement transitions in symmetric random quantum circuits},\ }\href
  {https://doi.org/10.1038/s41567-020-01112-z} {\bibfield  {journal} {\bibinfo
  {journal} {Nat. Phys.}\ }\textbf {\bibinfo {volume} {17}},\ \bibinfo {pages}
  {342} (\bibinfo {year} {2021})}\BibitemShut {NoStop}%
\bibitem [{\citenamefont {Sang}\ and\ \citenamefont
  {Hsieh}(2021)}]{Sang2021measurement}%
  \BibitemOpen
  \bibfield  {author} {\bibinfo {author} {\bibfnamefont {S.}~\bibnamefont
  {Sang}}\ and\ \bibinfo {author} {\bibfnamefont {T.~H.}\ \bibnamefont
  {Hsieh}},\ }\bibfield  {title} {\bibinfo {title} {Measurement-protected
  quantum phases},\ }\href {https://doi.org/10.1103/PhysRevResearch.3.023200}
  {\bibfield  {journal} {\bibinfo  {journal} {Phys. Rev. Research}\ }\textbf
  {\bibinfo {volume} {3}},\ \bibinfo {pages} {023200} (\bibinfo {year}
  {2021})}\BibitemShut {NoStop}%
\bibitem [{\citenamefont {Zhang}\ \emph {et~al.}(2020)\citenamefont {Zhang},
  \citenamefont {Reyes}, \citenamefont {Kourtis}, \citenamefont {Chamon},
  \citenamefont {Mucciolo},\ and\ \citenamefont {Ruckenstein}}]{Zhang2020}%
  \BibitemOpen
  \bibfield  {author} {\bibinfo {author} {\bibfnamefont {L.}~\bibnamefont
  {Zhang}}, \bibinfo {author} {\bibfnamefont {J.~A.}\ \bibnamefont {Reyes}},
  \bibinfo {author} {\bibfnamefont {S.}~\bibnamefont {Kourtis}}, \bibinfo
  {author} {\bibfnamefont {C.}~\bibnamefont {Chamon}}, \bibinfo {author}
  {\bibfnamefont {E.~R.}\ \bibnamefont {Mucciolo}},\ and\ \bibinfo {author}
  {\bibfnamefont {A.~E.}\ \bibnamefont {Ruckenstein}},\ }\bibfield  {title}
  {\bibinfo {title} {Nonuniversal entanglement level statistics in
  projection-driven quantum circuits},\ }\href
  {https://doi.org/10.1103/PhysRevB.101.235104} {\bibfield  {journal} {\bibinfo
   {journal} {Phys. Rev. B}\ }\textbf {\bibinfo {volume} {101}},\ \bibinfo
  {pages} {235104} (\bibinfo {year} {2020})}\BibitemShut {NoStop}%
\bibitem [{\citenamefont {Choi}\ \emph {et~al.}(2020)\citenamefont {Choi},
  \citenamefont {Bao}, \citenamefont {Qi},\ and\ \citenamefont
  {Altman}}]{Choi2020}%
  \BibitemOpen
  \bibfield  {author} {\bibinfo {author} {\bibfnamefont {S.}~\bibnamefont
  {Choi}}, \bibinfo {author} {\bibfnamefont {Y.}~\bibnamefont {Bao}}, \bibinfo
  {author} {\bibfnamefont {X.-L.}\ \bibnamefont {Qi}},\ and\ \bibinfo {author}
  {\bibfnamefont {E.}~\bibnamefont {Altman}},\ }\bibfield  {title} {\bibinfo
  {title} {Quantum error correction in scrambling dynamics and
  measurement-induced phase transition},\ }\href
  {https://doi.org/10.1103/PhysRevLett.125.030505} {\bibfield  {journal}
  {\bibinfo  {journal} {Phys. Rev. Lett.}\ }\textbf {\bibinfo {volume} {125}},\
  \bibinfo {pages} {030505} (\bibinfo {year} {2020})}\BibitemShut {NoStop}%
\bibitem [{\citenamefont {Turkeshi}\ \emph {et~al.}(2020)\citenamefont
  {Turkeshi}, \citenamefont {Fazio},\ and\ \citenamefont
  {Dalmonte}}]{Turkeshi2020}%
  \BibitemOpen
  \bibfield  {author} {\bibinfo {author} {\bibfnamefont {X.}~\bibnamefont
  {Turkeshi}}, \bibinfo {author} {\bibfnamefont {R.}~\bibnamefont {Fazio}},\
  and\ \bibinfo {author} {\bibfnamefont {M.}~\bibnamefont {Dalmonte}},\
  }\bibfield  {title} {\bibinfo {title} {Measurement-induced criticality in
  $(2+1)$-dimensional hybrid quantum circuits},\ }\href
  {https://doi.org/10.1103/PhysRevB.102.014315} {\bibfield  {journal} {\bibinfo
   {journal} {Phys. Rev. B}\ }\textbf {\bibinfo {volume} {102}},\ \bibinfo
  {pages} {014315} (\bibinfo {year} {2020})}\BibitemShut {NoStop}%
\bibitem [{\citenamefont {Gullans}\ and\ \citenamefont
  {Huse}(2020{\natexlab{b}})}]{Gullans2020}%
  \BibitemOpen
  \bibfield  {author} {\bibinfo {author} {\bibfnamefont {M.~J.}\ \bibnamefont
  {Gullans}}\ and\ \bibinfo {author} {\bibfnamefont {D.~A.}\ \bibnamefont
  {Huse}},\ }\bibfield  {title} {\bibinfo {title} {Scalable probes of
  measurement-induced criticality},\ }\href
  {https://doi.org/10.1103/PhysRevLett.125.070606} {\bibfield  {journal}
  {\bibinfo  {journal} {Phys. Rev. Lett.}\ }\textbf {\bibinfo {volume} {125}},\
  \bibinfo {pages} {070606} (\bibinfo {year} {2020}{\natexlab{b}})}\BibitemShut
  {NoStop}%
\bibitem [{\citenamefont {Nahum}\ \emph {et~al.}(2021)\citenamefont {Nahum},
  \citenamefont {Roy}, \citenamefont {Skinner},\ and\ \citenamefont
  {Ruhman}}]{Nahum2021}%
  \BibitemOpen
  \bibfield  {author} {\bibinfo {author} {\bibfnamefont {A.}~\bibnamefont
  {Nahum}}, \bibinfo {author} {\bibfnamefont {S.}~\bibnamefont {Roy}}, \bibinfo
  {author} {\bibfnamefont {B.}~\bibnamefont {Skinner}},\ and\ \bibinfo {author}
  {\bibfnamefont {J.}~\bibnamefont {Ruhman}},\ }\bibfield  {title} {\bibinfo
  {title} {Measurement and entanglement phase transitions in all-to-all quantum
  circuits, on quantum trees, and in {Landau-Ginsburg} theory},\ }\href
  {https://doi.org/10.1103/PRXQuantum.2.010352} {\bibfield  {journal} {\bibinfo
   {journal} {PRX Quantum}\ }\textbf {\bibinfo {volume} {2}},\ \bibinfo {pages}
  {010352} (\bibinfo {year} {2021})}\BibitemShut {NoStop}%
\bibitem [{\citenamefont {Cao}\ \emph {et~al.}(2019)\citenamefont {Cao},
  \citenamefont {Tilloy},\ and\ \citenamefont {Luca}}]{Cao2019}%
  \BibitemOpen
  \bibfield  {author} {\bibinfo {author} {\bibfnamefont {X.}~\bibnamefont
  {Cao}}, \bibinfo {author} {\bibfnamefont {A.}~\bibnamefont {Tilloy}},\ and\
  \bibinfo {author} {\bibfnamefont {A.~D.}\ \bibnamefont {Luca}},\ }\bibfield
  {title} {\bibinfo {title} {{Entanglement in a fermion chain under continuous
  monitoring}},\ }\href {https://doi.org/10.21468/SciPostPhys.7.2.024}
  {\bibfield  {journal} {\bibinfo  {journal} {SciPost Phys.}\ }\textbf
  {\bibinfo {volume} {7}},\ \bibinfo {pages} {24} (\bibinfo {year}
  {2019})}\BibitemShut {NoStop}%
\bibitem [{\citenamefont {Alberton}\ \emph {et~al.}(2021)\citenamefont
  {Alberton}, \citenamefont {Buchhold},\ and\ \citenamefont
  {Diehl}}]{Alberton2021}%
  \BibitemOpen
  \bibfield  {author} {\bibinfo {author} {\bibfnamefont {O.}~\bibnamefont
  {Alberton}}, \bibinfo {author} {\bibfnamefont {M.}~\bibnamefont {Buchhold}},\
  and\ \bibinfo {author} {\bibfnamefont {S.}~\bibnamefont {Diehl}},\ }\bibfield
   {title} {\bibinfo {title} {Entanglement transition in a monitored
  free-fermion chain: From extended criticality to area law},\ }\href
  {https://doi.org/10.1103/PhysRevLett.126.170602} {\bibfield  {journal}
  {\bibinfo  {journal} {Phys. Rev. Lett.}\ }\textbf {\bibinfo {volume} {126}},\
  \bibinfo {pages} {170602} (\bibinfo {year} {2021})}\BibitemShut {NoStop}%
\bibitem [{\citenamefont {Tang}\ and\ \citenamefont {Zhu}(2020)}]{Tang2020}%
  \BibitemOpen
  \bibfield  {author} {\bibinfo {author} {\bibfnamefont {Q.}~\bibnamefont
  {Tang}}\ and\ \bibinfo {author} {\bibfnamefont {W.}~\bibnamefont {Zhu}},\
  }\bibfield  {title} {\bibinfo {title} {Measurement-induced phase transition:
  A case study in the nonintegrable model by density-matrix renormalization
  group calculations},\ }\href
  {https://doi.org/10.1103/PhysRevResearch.2.013022} {\bibfield  {journal}
  {\bibinfo  {journal} {Phys. Rev. Research}\ }\textbf {\bibinfo {volume}
  {2}},\ \bibinfo {pages} {013022} (\bibinfo {year} {2020})}\BibitemShut
  {NoStop}%
\bibitem [{\citenamefont {Goto}\ and\ \citenamefont
  {Danshita}(2020)}]{Goto2020}%
  \BibitemOpen
  \bibfield  {author} {\bibinfo {author} {\bibfnamefont {S.}~\bibnamefont
  {Goto}}\ and\ \bibinfo {author} {\bibfnamefont {I.}~\bibnamefont
  {Danshita}},\ }\bibfield  {title} {\bibinfo {title} {Measurement-induced
  transitions of the entanglement scaling law in ultracold gases with
  controllable dissipation},\ }\href
  {https://doi.org/10.1103/PhysRevA.102.033316} {\bibfield  {journal} {\bibinfo
   {journal} {Phys. Rev. A}\ }\textbf {\bibinfo {volume} {102}},\ \bibinfo
  {pages} {033316} (\bibinfo {year} {2020})}\BibitemShut {NoStop}%
\bibitem [{\citenamefont {Fuji}\ and\ \citenamefont {Ashida}(2020)}]{Fuji2020}%
  \BibitemOpen
  \bibfield  {author} {\bibinfo {author} {\bibfnamefont {Y.}~\bibnamefont
  {Fuji}}\ and\ \bibinfo {author} {\bibfnamefont {Y.}~\bibnamefont {Ashida}},\
  }\bibfield  {title} {\bibinfo {title} {Measurement-induced quantum
  criticality under continuous monitoring},\ }\href
  {https://doi.org/10.1103/PhysRevB.102.054302} {\bibfield  {journal} {\bibinfo
   {journal} {Phys. Rev. B}\ }\textbf {\bibinfo {volume} {102}},\ \bibinfo
  {pages} {054302} (\bibinfo {year} {2020})}\BibitemShut {NoStop}%
\bibitem [{\citenamefont {Rossini}\ and\ \citenamefont
  {Vicari}(2020)}]{Rossini2020}%
  \BibitemOpen
  \bibfield  {author} {\bibinfo {author} {\bibfnamefont {D.}~\bibnamefont
  {Rossini}}\ and\ \bibinfo {author} {\bibfnamefont {E.}~\bibnamefont
  {Vicari}},\ }\bibfield  {title} {\bibinfo {title} {Measurement-induced
  dynamics of many-body systems at quantum criticality},\ }\href
  {https://doi.org/10.1103/PhysRevB.102.035119} {\bibfield  {journal} {\bibinfo
   {journal} {Phys. Rev. B}\ }\textbf {\bibinfo {volume} {102}},\ \bibinfo
  {pages} {035119} (\bibinfo {year} {2020})}\BibitemShut {NoStop}%
\bibitem [{\citenamefont {Lunt}\ and\ \citenamefont {Pal}(2020)}]{Lunt2020}%
  \BibitemOpen
  \bibfield  {author} {\bibinfo {author} {\bibfnamefont {O.}~\bibnamefont
  {Lunt}}\ and\ \bibinfo {author} {\bibfnamefont {A.}~\bibnamefont {Pal}},\
  }\bibfield  {title} {\bibinfo {title} {Measurement-induced entanglement
  transitions in many-body localized systems},\ }\href
  {https://doi.org/10.1103/PhysRevResearch.2.043072} {\bibfield  {journal}
  {\bibinfo  {journal} {Phys. Rev. Research}\ }\textbf {\bibinfo {volume}
  {2}},\ \bibinfo {pages} {043072} (\bibinfo {year} {2020})}\BibitemShut
  {NoStop}%
\bibitem [{\citenamefont {Chen}\ \emph {et~al.}(2020)\citenamefont {Chen},
  \citenamefont {Li}, \citenamefont {Fisher},\ and\ \citenamefont
  {Lucas}}]{Chen2020}%
  \BibitemOpen
  \bibfield  {author} {\bibinfo {author} {\bibfnamefont {X.}~\bibnamefont
  {Chen}}, \bibinfo {author} {\bibfnamefont {Y.}~\bibnamefont {Li}}, \bibinfo
  {author} {\bibfnamefont {M.~P.~A.}\ \bibnamefont {Fisher}},\ and\ \bibinfo
  {author} {\bibfnamefont {A.}~\bibnamefont {Lucas}},\ }\bibfield  {title}
  {\bibinfo {title} {Emergent conformal symmetry in nonunitary random dynamics
  of free fermions},\ }\href {https://doi.org/10.1103/PhysRevResearch.2.033017}
  {\bibfield  {journal} {\bibinfo  {journal} {Phys. Rev. Research}\ }\textbf
  {\bibinfo {volume} {2}},\ \bibinfo {pages} {033017} (\bibinfo {year}
  {2020})}\BibitemShut {NoStop}%
\bibitem [{\citenamefont {Liu}\ \emph {et~al.}(2021)\citenamefont {Liu},
  \citenamefont {Zhang},\ and\ \citenamefont {Chen}}]{Liu2021}%
  \BibitemOpen
  \bibfield  {author} {\bibinfo {author} {\bibfnamefont {C.}~\bibnamefont
  {Liu}}, \bibinfo {author} {\bibfnamefont {P.}~\bibnamefont {Zhang}},\ and\
  \bibinfo {author} {\bibfnamefont {X.}~\bibnamefont {Chen}},\ }\bibfield
  {title} {\bibinfo {title} {{Non-unitary dynamics of Sachdev-Ye-Kitaev
  chain}},\ }\href {https://doi.org/10.21468/SciPostPhys.10.2.048} {\bibfield
  {journal} {\bibinfo  {journal} {SciPost Phys.}\ }\textbf {\bibinfo {volume}
  {10}},\ \bibinfo {pages} {48} (\bibinfo {year} {2021})}\BibitemShut {NoStop}%
\bibitem [{\citenamefont {Biella}\ and\ \citenamefont
  {Schir\'o}(2020)}]{Biella2020}%
  \BibitemOpen
  \bibfield  {author} {\bibinfo {author} {\bibfnamefont {A.}~\bibnamefont
  {Biella}}\ and\ \bibinfo {author} {\bibfnamefont {M.}~\bibnamefont
  {Schir\'o}},\ }\bibfield  {title} {\bibinfo {title} {Many-body quantum zeno
  effect and measurement-induced subradiance transition},\ }\Eprint
  {https://arxiv.org/abs/2011.11620} {arXiv:2011.11620}  (\bibinfo {year}
  {2020})\BibitemShut {NoStop}%
\bibitem [{\citenamefont {Gopalakrishnan}\ and\ \citenamefont
  {Gullans}(2021)}]{Gopalakrishnan2021}%
  \BibitemOpen
  \bibfield  {author} {\bibinfo {author} {\bibfnamefont {S.}~\bibnamefont
  {Gopalakrishnan}}\ and\ \bibinfo {author} {\bibfnamefont {M.~J.}\
  \bibnamefont {Gullans}},\ }\bibfield  {title} {\bibinfo {title} {Entanglement
  and purification transitions in non-{Hermitian} quantum mechanics},\ }\href
  {https://doi.org/10.1103/PhysRevLett.126.170503} {\bibfield  {journal}
  {\bibinfo  {journal} {Phys. Rev. Lett.}\ }\textbf {\bibinfo {volume} {126}},\
  \bibinfo {pages} {170503} (\bibinfo {year} {2021})}\BibitemShut {NoStop}%
\bibitem [{\citenamefont {Jian}\ \emph {et~al.}(2021)\citenamefont {Jian},
  \citenamefont {Yang}, \citenamefont {Bi},\ and\ \citenamefont
  {Chen}}]{Jian2021}%
  \BibitemOpen
  \bibfield  {author} {\bibinfo {author} {\bibfnamefont {S.-K.}\ \bibnamefont
  {Jian}}, \bibinfo {author} {\bibfnamefont {Z.-C.}\ \bibnamefont {Yang}},
  \bibinfo {author} {\bibfnamefont {Z.}~\bibnamefont {Bi}},\ and\ \bibinfo
  {author} {\bibfnamefont {X.}~\bibnamefont {Chen}},\ }\bibfield  {title}
  {\bibinfo {title} {{Yang}-lee edge singularity triggered entanglement
  transition},\ }\Eprint {https://arxiv.org/abs/2101.04115} {arXiv:2101.04115}
  (\bibinfo {year} {2021})\BibitemShut {NoStop}%
\bibitem [{\citenamefont {Tang}\ \emph {et~al.}(2021)\citenamefont {Tang},
  \citenamefont {Chen},\ and\ \citenamefont {Zhu}}]{Tang2021}%
  \BibitemOpen
  \bibfield  {author} {\bibinfo {author} {\bibfnamefont {Q.}~\bibnamefont
  {Tang}}, \bibinfo {author} {\bibfnamefont {X.}~\bibnamefont {Chen}},\ and\
  \bibinfo {author} {\bibfnamefont {W.}~\bibnamefont {Zhu}},\ }\bibfield
  {title} {\bibinfo {title} {Quantum criticality in the nonunitary dynamics of
  $(2+1)$-dimensional free fermions},\ }\href
  {https://doi.org/10.1103/PhysRevB.103.174303} {\bibfield  {journal} {\bibinfo
   {journal} {Phys. Rev. B}\ }\textbf {\bibinfo {volume} {103}},\ \bibinfo
  {pages} {174303} (\bibinfo {year} {2021})}\BibitemShut {NoStop}%
\bibitem [{\citenamefont {Turkeshi}(2021)}]{Turkeshi2021}%
  \BibitemOpen
  \bibfield  {author} {\bibinfo {author} {\bibfnamefont {X.}~\bibnamefont
  {Turkeshi}},\ }\bibfield  {title} {\bibinfo {title} {Measurement-induced
  criticality as a data-structure transition},\ }\Eprint
  {https://arxiv.org/abs/2101.06245} {arXiv:2101.06245}  (\bibinfo {year}
  {2021})\BibitemShut {NoStop}%
\bibitem [{\citenamefont {Buchhold}\ \emph {et~al.}(2021)\citenamefont
  {Buchhold}, \citenamefont {Minoguchi}, \citenamefont {Altland},\ and\
  \citenamefont {Diehl}}]{Buchhold2021}%
  \BibitemOpen
  \bibfield  {author} {\bibinfo {author} {\bibfnamefont {M.}~\bibnamefont
  {Buchhold}}, \bibinfo {author} {\bibfnamefont {Y.}~\bibnamefont {Minoguchi}},
  \bibinfo {author} {\bibfnamefont {A.}~\bibnamefont {Altland}},\ and\ \bibinfo
  {author} {\bibfnamefont {S.}~\bibnamefont {Diehl}},\ }\bibfield  {title}
  {\bibinfo {title} {Effective theory for the measurement-induced phase
  transition of dirac fermions},\ }\Eprint {https://arxiv.org/abs/2102.08381}
  {arXiv:2102.08381}  (\bibinfo {year} {2021})\BibitemShut {NoStop}%
\bibitem [{\citenamefont {Lang}\ and\ \citenamefont
  {B\"uchler}(2020)}]{Lang2020}%
  \BibitemOpen
  \bibfield  {author} {\bibinfo {author} {\bibfnamefont {N.}~\bibnamefont
  {Lang}}\ and\ \bibinfo {author} {\bibfnamefont {H.~P.}\ \bibnamefont
  {B\"uchler}},\ }\bibfield  {title} {\bibinfo {title} {Entanglement transition
  in the projective transverse field {Ising} model},\ }\href
  {https://doi.org/10.1103/PhysRevB.102.094204} {\bibfield  {journal} {\bibinfo
   {journal} {Phys. Rev. B}\ }\textbf {\bibinfo {volume} {102}},\ \bibinfo
  {pages} {094204} (\bibinfo {year} {2020})}\BibitemShut {NoStop}%
\bibitem [{\citenamefont {Ippoliti}\ \emph {et~al.}(2021)\citenamefont
  {Ippoliti}, \citenamefont {Gullans}, \citenamefont {Gopalakrishnan},
  \citenamefont {Huse},\ and\ \citenamefont
  {Khemani}}]{Ippoliti2021entanglement}%
  \BibitemOpen
  \bibfield  {author} {\bibinfo {author} {\bibfnamefont {M.}~\bibnamefont
  {Ippoliti}}, \bibinfo {author} {\bibfnamefont {M.~J.}\ \bibnamefont
  {Gullans}}, \bibinfo {author} {\bibfnamefont {S.}~\bibnamefont
  {Gopalakrishnan}}, \bibinfo {author} {\bibfnamefont {D.~A.}\ \bibnamefont
  {Huse}},\ and\ \bibinfo {author} {\bibfnamefont {V.}~\bibnamefont
  {Khemani}},\ }\bibfield  {title} {\bibinfo {title} {Entanglement phase
  transitions in measurement-only dynamics},\ }\href
  {https://doi.org/10.1103/PhysRevX.11.011030} {\bibfield  {journal} {\bibinfo
  {journal} {Phys. Rev. X}\ }\textbf {\bibinfo {volume} {11}},\ \bibinfo
  {pages} {011030} (\bibinfo {year} {2021})}\BibitemShut {NoStop}%
\bibitem [{\citenamefont {Van~Regemortel}\ \emph {et~al.}(2021)\citenamefont
  {Van~Regemortel}, \citenamefont {Cian}, \citenamefont {Seif}, \citenamefont
  {Dehghani},\ and\ \citenamefont {Hafezi}}]{VanRegemortel2021}%
  \BibitemOpen
  \bibfield  {author} {\bibinfo {author} {\bibfnamefont {M.}~\bibnamefont
  {Van~Regemortel}}, \bibinfo {author} {\bibfnamefont {Z.-P.}\ \bibnamefont
  {Cian}}, \bibinfo {author} {\bibfnamefont {A.}~\bibnamefont {Seif}}, \bibinfo
  {author} {\bibfnamefont {H.}~\bibnamefont {Dehghani}},\ and\ \bibinfo
  {author} {\bibfnamefont {M.}~\bibnamefont {Hafezi}},\ }\bibfield  {title}
  {\bibinfo {title} {Entanglement entropy scaling transition under competing
  monitoring protocols},\ }\href
  {https://doi.org/10.1103/PhysRevLett.126.123604} {\bibfield  {journal}
  {\bibinfo  {journal} {Phys. Rev. Lett.}\ }\textbf {\bibinfo {volume} {126}},\
  \bibinfo {pages} {123604} (\bibinfo {year} {2021})}\BibitemShut {NoStop}%
\bibitem [{\citenamefont {Vijay}(2020)}]{Vijay2020}%
  \BibitemOpen
  \bibfield  {author} {\bibinfo {author} {\bibfnamefont {S.}~\bibnamefont
  {Vijay}},\ }\bibfield  {title} {\bibinfo {title} {Measurement-driven phase
  transition within a volume-law entangled phase},\ }\Eprint
  {https://arxiv.org/abs/2005.03052} {arXiv:2005.03052}  (\bibinfo {year}
  {2020})\BibitemShut {NoStop}%
\bibitem [{\citenamefont {Nahum}\ and\ \citenamefont
  {Skinner}(2020)}]{Nahum2020defects}%
  \BibitemOpen
  \bibfield  {author} {\bibinfo {author} {\bibfnamefont {A.}~\bibnamefont
  {Nahum}}\ and\ \bibinfo {author} {\bibfnamefont {B.}~\bibnamefont
  {Skinner}},\ }\bibfield  {title} {\bibinfo {title} {Entanglement and dynamics
  of diffusion-annihilation processes with {{Majorana}} defects},\ }\href
  {https://doi.org/10.1103/PhysRevResearch.2.023288} {\bibfield  {journal}
  {\bibinfo  {journal} {Phys. Rev. Research}\ }\textbf {\bibinfo {volume}
  {2}},\ \bibinfo {pages} {023288} (\bibinfo {year} {2020})}\BibitemShut
  {NoStop}%
\bibitem [{\citenamefont {Li}\ and\ \citenamefont {Fisher}(2021)}]{Li2021}%
  \BibitemOpen
  \bibfield  {author} {\bibinfo {author} {\bibfnamefont {Y.}~\bibnamefont
  {Li}}\ and\ \bibinfo {author} {\bibfnamefont {M.~P.~A.}\ \bibnamefont
  {Fisher}},\ }\bibfield  {title} {\bibinfo {title} {Statistical mechanics of
  quantum error correcting codes},\ }\href
  {https://doi.org/10.1103/PhysRevB.103.104306} {\bibfield  {journal} {\bibinfo
   {journal} {Phys. Rev. B}\ }\textbf {\bibinfo {volume} {103}},\ \bibinfo
  {pages} {104306} (\bibinfo {year} {2021})}\BibitemShut {NoStop}%
\bibitem [{\citenamefont {Lunt}\ \emph {et~al.}(2020)\citenamefont {Lunt},
  \citenamefont {Szyniszewski},\ and\ \citenamefont {Pal}}]{Lunt2020hybridity}%
  \BibitemOpen
  \bibfield  {author} {\bibinfo {author} {\bibfnamefont {O.}~\bibnamefont
  {Lunt}}, \bibinfo {author} {\bibfnamefont {M.}~\bibnamefont {Szyniszewski}},\
  and\ \bibinfo {author} {\bibfnamefont {A.}~\bibnamefont {Pal}},\ }\bibfield
  {title} {\bibinfo {title} {Dimensional hybridity in measurement-induced
  criticality},\ }\Eprint {https://arxiv.org/abs/2012.03857} {arXiv:2012.03857}
   (\bibinfo {year} {2020})\BibitemShut {NoStop}%
\bibitem [{\citenamefont {Gullans}\ \emph {et~al.}(2020)\citenamefont
  {Gullans}, \citenamefont {Krastanov}, \citenamefont {Huse}, \citenamefont
  {Jiang},\ and\ \citenamefont {Flammia}}]{Gullans2020lowdepth}%
  \BibitemOpen
  \bibfield  {author} {\bibinfo {author} {\bibfnamefont {M.~J.}\ \bibnamefont
  {Gullans}}, \bibinfo {author} {\bibfnamefont {S.}~\bibnamefont {Krastanov}},
  \bibinfo {author} {\bibfnamefont {D.~A.}\ \bibnamefont {Huse}}, \bibinfo
  {author} {\bibfnamefont {L.}~\bibnamefont {Jiang}},\ and\ \bibinfo {author}
  {\bibfnamefont {S.~T.}\ \bibnamefont {Flammia}},\ }\bibfield  {title}
  {\bibinfo {title} {Quantum coding with low-depth random circuits},\ }\Eprint
  {https://arxiv.org/abs/2010.09775} {arXiv:2010.09775}  (\bibinfo {year}
  {2020})\BibitemShut {NoStop}%
\bibitem [{\citenamefont {Fidkowski}\ \emph {et~al.}(2021)\citenamefont
  {Fidkowski}, \citenamefont {Haah},\ and\ \citenamefont
  {Hastings}}]{Fidkowski2021}%
  \BibitemOpen
  \bibfield  {author} {\bibinfo {author} {\bibfnamefont {L.}~\bibnamefont
  {Fidkowski}}, \bibinfo {author} {\bibfnamefont {J.}~\bibnamefont {Haah}},\
  and\ \bibinfo {author} {\bibfnamefont {M.~B.}\ \bibnamefont {Hastings}},\
  }\bibfield  {title} {\bibinfo {title} {How dynamical quantum memories
  forget},\ }\href {https://doi.org/10.22331/q-2021-01-17-382} {\bibfield
  {journal} {\bibinfo  {journal} {{Quantum}}\ }\textbf {\bibinfo {volume}
  {5}},\ \bibinfo {pages} {382} (\bibinfo {year} {2021})}\BibitemShut {NoStop}%
\bibitem [{\citenamefont {Maimbourg}\ \emph {et~al.}(2021)\citenamefont
  {Maimbourg}, \citenamefont {Basko}, \citenamefont {Holzmann},\ and\
  \citenamefont {Rosso}}]{Maimbourg2021}%
  \BibitemOpen
  \bibfield  {author} {\bibinfo {author} {\bibfnamefont {T.}~\bibnamefont
  {Maimbourg}}, \bibinfo {author} {\bibfnamefont {D.~M.}\ \bibnamefont
  {Basko}}, \bibinfo {author} {\bibfnamefont {M.}~\bibnamefont {Holzmann}},\
  and\ \bibinfo {author} {\bibfnamefont {A.}~\bibnamefont {Rosso}},\ }\bibfield
   {title} {\bibinfo {title} {Bath-induced zeno localization in driven
  many-body quantum systems},\ }\href
  {https://doi.org/10.1103/PhysRevLett.126.120603} {\bibfield  {journal}
  {\bibinfo  {journal} {Phys. Rev. Lett.}\ }\textbf {\bibinfo {volume} {126}},\
  \bibinfo {pages} {120603} (\bibinfo {year} {2021})}\BibitemShut {NoStop}%
\bibitem [{\citenamefont {Iaconis}\ \emph {et~al.}(2020)\citenamefont
  {Iaconis}, \citenamefont {Lucas},\ and\ \citenamefont {Chen}}]{Iaconis2020}%
  \BibitemOpen
  \bibfield  {author} {\bibinfo {author} {\bibfnamefont {J.}~\bibnamefont
  {Iaconis}}, \bibinfo {author} {\bibfnamefont {A.}~\bibnamefont {Lucas}},\
  and\ \bibinfo {author} {\bibfnamefont {X.}~\bibnamefont {Chen}},\ }\bibfield
  {title} {\bibinfo {title} {Measurement-induced phase transitions in quantum
  automaton circuits},\ }\href {https://doi.org/10.1103/PhysRevB.102.224311}
  {\bibfield  {journal} {\bibinfo  {journal} {Phys. Rev. B}\ }\textbf {\bibinfo
  {volume} {102}},\ \bibinfo {pages} {224311} (\bibinfo {year}
  {2020})}\BibitemShut {NoStop}%
\bibitem [{\citenamefont {Ippoliti}\ and\ \citenamefont
  {Khemani}(2021)}]{Ippoliti2021postselection}%
  \BibitemOpen
  \bibfield  {author} {\bibinfo {author} {\bibfnamefont {M.}~\bibnamefont
  {Ippoliti}}\ and\ \bibinfo {author} {\bibfnamefont {V.}~\bibnamefont
  {Khemani}},\ }\bibfield  {title} {\bibinfo {title} {Postselection-free
  entanglement dynamics via spacetime duality},\ }\href
  {https://doi.org/10.1103/PhysRevLett.126.060501} {\bibfield  {journal}
  {\bibinfo  {journal} {Phys. Rev. Lett.}\ }\textbf {\bibinfo {volume} {126}},\
  \bibinfo {pages} {060501} (\bibinfo {year} {2021})}\BibitemShut {NoStop}%
\bibitem [{\citenamefont {Lavasani}\ \emph {et~al.}(2020)\citenamefont
  {Lavasani}, \citenamefont {Alavirad},\ and\ \citenamefont
  {Barkeshli}}]{Lavasani2020topological}%
  \BibitemOpen
  \bibfield  {author} {\bibinfo {author} {\bibfnamefont {A.}~\bibnamefont
  {Lavasani}}, \bibinfo {author} {\bibfnamefont {Y.}~\bibnamefont {Alavirad}},\
  and\ \bibinfo {author} {\bibfnamefont {M.}~\bibnamefont {Barkeshli}},\
  }\bibfield  {title} {\bibinfo {title} {Topological order and criticality in
  (2+1){{D}} monitored random quantum circuits},\ }\Eprint
  {https://arxiv.org/abs/2011.06595} {arXiv:2011.06595}  (\bibinfo {year}
  {2020})\BibitemShut {NoStop}%
\bibitem [{\citenamefont {Sang}\ \emph {et~al.}(2020)\citenamefont {Sang},
  \citenamefont {Li}, \citenamefont {Zhou}, \citenamefont {Chen}, \citenamefont
  {Hsieh},\ and\ \citenamefont {Fisher}}]{Sang2020entanglement}%
  \BibitemOpen
  \bibfield  {author} {\bibinfo {author} {\bibfnamefont {S.}~\bibnamefont
  {Sang}}, \bibinfo {author} {\bibfnamefont {Y.}~\bibnamefont {Li}}, \bibinfo
  {author} {\bibfnamefont {T.}~\bibnamefont {Zhou}}, \bibinfo {author}
  {\bibfnamefont {X.}~\bibnamefont {Chen}}, \bibinfo {author} {\bibfnamefont
  {T.~H.}\ \bibnamefont {Hsieh}},\ and\ \bibinfo {author} {\bibfnamefont
  {M.~P.~A.}\ \bibnamefont {Fisher}},\ }\bibfield  {title} {\bibinfo {title}
  {Entanglement negativity at measurement-induced criticality},\ }\Eprint
  {https://arxiv.org/abs/2012.00031} {arXiv:2012.00031}  (\bibinfo {year}
  {2020})\BibitemShut {NoStop}%
\bibitem [{\citenamefont {Shi}\ \emph {et~al.}(2020)\citenamefont {Shi},
  \citenamefont {Dai},\ and\ \citenamefont {Lu}}]{Shi2020}%
  \BibitemOpen
  \bibfield  {author} {\bibinfo {author} {\bibfnamefont {B.}~\bibnamefont
  {Shi}}, \bibinfo {author} {\bibfnamefont {X.}~\bibnamefont {Dai}},\ and\
  \bibinfo {author} {\bibfnamefont {Y.-M.}\ \bibnamefont {Lu}},\ }\bibfield
  {title} {\bibinfo {title} {Entanglement negativity at the critical point of
  measurement-driven transition},\ }\Eprint {https://arxiv.org/abs/2012.00040}
  {arXiv:2012.00040}  (\bibinfo {year} {2020})\BibitemShut {NoStop}%
\bibitem [{\citenamefont {Bao}\ \emph {et~al.}(2021)\citenamefont {Bao},
  \citenamefont {Choi},\ and\ \citenamefont {Altman}}]{Bao2021}%
  \BibitemOpen
  \bibfield  {author} {\bibinfo {author} {\bibfnamefont {Y.}~\bibnamefont
  {Bao}}, \bibinfo {author} {\bibfnamefont {S.}~\bibnamefont {Choi}},\ and\
  \bibinfo {author} {\bibfnamefont {E.}~\bibnamefont {Altman}},\ }\bibfield
  {title} {\bibinfo {title} {Symmetry enriched phases of quantum circuits},\
  }\Eprint {https://arxiv.org/abs/2102.09164} {arXiv:2102.09164}  (\bibinfo
  {year} {2021})\BibitemShut {NoStop}%
\bibitem [{\citenamefont {Rossini}\ and\ \citenamefont
  {Vicari}(2021)}]{Rossini2021}%
  \BibitemOpen
  \bibfield  {author} {\bibinfo {author} {\bibfnamefont {D.}~\bibnamefont
  {Rossini}}\ and\ \bibinfo {author} {\bibfnamefont {E.}~\bibnamefont
  {Vicari}},\ }\bibfield  {title} {\bibinfo {title} {Coherent and dissipative
  dynamics at quantum phase transitions},\ }\Eprint
  {https://arxiv.org/abs/2103.02626} {arXiv:2103.02626}  (\bibinfo {year}
  {2021})\BibitemShut {NoStop}%
\bibitem [{\citenamefont {Tsomokos}\ \emph {et~al.}(2007)\citenamefont
  {Tsomokos}, \citenamefont {Hartmann}, \citenamefont {Huelga},\ and\
  \citenamefont {Plenio}}]{tsomokos2007}%
  \BibitemOpen
  \bibfield  {author} {\bibinfo {author} {\bibfnamefont {D.~I.}\ \bibnamefont
  {Tsomokos}}, \bibinfo {author} {\bibfnamefont {M.~J.}\ \bibnamefont
  {Hartmann}}, \bibinfo {author} {\bibfnamefont {S.~F.}\ \bibnamefont
  {Huelga}},\ and\ \bibinfo {author} {\bibfnamefont {M.~B.}\ \bibnamefont
  {Plenio}},\ }\bibfield  {title} {\bibinfo {title} {Entanglement dynamics in
  chains of qubits with noise and disorder},\ }\href
  {https://doi.org/10.1088/1367-2630/9/3/079} {\bibfield  {journal} {\bibinfo
  {journal} {New J. Physics}\ }\textbf {\bibinfo {volume} {9}},\ \bibinfo
  {pages} {79} (\bibinfo {year} {2007})}\BibitemShut {NoStop}%
\bibitem [{\citenamefont {Knap}(2018)}]{knap2018}%
  \BibitemOpen
  \bibfield  {author} {\bibinfo {author} {\bibfnamefont {M.}~\bibnamefont
  {Knap}},\ }\bibfield  {title} {\bibinfo {title} {Entanglement production and
  information scrambling in a noisy spin system},\ }\href
  {https://doi.org/10.1103/PhysRevB.98.184416} {\bibfield  {journal} {\bibinfo
  {journal} {Phys. Rev. B}\ }\textbf {\bibinfo {volume} {98}},\ \bibinfo
  {pages} {184416} (\bibinfo {year} {2018})}\BibitemShut {NoStop}%
\bibitem [{\citenamefont {Ghosh}\ and\ \citenamefont {Das}(2021)}]{ghosh2021}%
  \BibitemOpen
  \bibfield  {author} {\bibinfo {author} {\bibfnamefont {R.}~\bibnamefont
  {Ghosh}}\ and\ \bibinfo {author} {\bibfnamefont {A.}~\bibnamefont {Das}},\
  }\bibfield  {title} {\bibinfo {title} {Disorder-induced enhancement of
  entanglement growth in one dimension: Information leakage at the scale of the
  localization length},\ }\href {https://doi.org/10.1103/PhysRevB.103.024202}
  {\bibfield  {journal} {\bibinfo  {journal} {Phys. Rev. B}\ }\textbf {\bibinfo
  {volume} {103}},\ \bibinfo {pages} {024202} (\bibinfo {year}
  {2021})}\BibitemShut {NoStop}%
\bibitem [{\citenamefont {Jacobs}(2014)}]{Jacobs2014quantum}%
  \BibitemOpen
  \bibfield  {author} {\bibinfo {author} {\bibfnamefont {K.}~\bibnamefont
  {Jacobs}},\ }\href {https://doi.org/10.1017/CBO9781139179027} {\emph
  {\bibinfo {title} {Quantum measurement theory and its applications}}}\
  (\bibinfo  {publisher} {Cambridge University Press},\ \bibinfo {year}
  {2014})\BibitemShut {NoStop}%
\bibitem [{\citenamefont {Wiseman}\ and\ \citenamefont
  {Milburn}(2009)}]{wiseman_milburn_2009}%
  \BibitemOpen
  \bibfield  {author} {\bibinfo {author} {\bibfnamefont {H.~M.}\ \bibnamefont
  {Wiseman}}\ and\ \bibinfo {author} {\bibfnamefont {G.~J.}\ \bibnamefont
  {Milburn}},\ }\href {https://doi.org/10.1017/CBO9780511813948} {\emph
  {\bibinfo {title} {Quantum Measurement and Control}}}\ (\bibinfo  {publisher}
  {Cambridge University Press},\ \bibinfo {year} {2009})\BibitemShut {NoStop}%
\bibitem [{\citenamefont {Hatano}\ and\ \citenamefont
  {Suzuki}(2005)}]{hatano2005finding}%
  \BibitemOpen
  \bibfield  {author} {\bibinfo {author} {\bibfnamefont {N.}~\bibnamefont
  {Hatano}}\ and\ \bibinfo {author} {\bibfnamefont {M.}~\bibnamefont
  {Suzuki}},\ }\bibfield  {title} {\bibinfo {title} {Finding exponential
  product formulas of higher orders},\ }in\ \href
  {https://doi.org/10.1007/11526216_2} {\emph {\bibinfo {booktitle} {Quantum
  annealing and other optimization methods}}}\ (\bibinfo  {publisher}
  {Springer},\ \bibinfo {year} {2005})\ pp.\ \bibinfo {pages}
  {37--68}\BibitemShut {NoStop}%
\bibitem [{\citenamefont {Zalka}(1998)}]{christof98Simulating}%
  \BibitemOpen
  \bibfield  {author} {\bibinfo {author} {\bibfnamefont {C.}~\bibnamefont
  {Zalka}},\ }\bibfield  {title} {\bibinfo {title} {Simulating quantum systems
  on a quantum computer},\ }\href {https://doi.org/10.1098/rspa.1998.0162}
  {\bibfield  {journal} {\bibinfo  {journal} {Proc. R. Soc. Lond. A.}\ }\textbf
  {\bibinfo {volume} {454}},\ \bibinfo {pages} {313} (\bibinfo {year}
  {1998})}\BibitemShut {NoStop}%
\bibitem [{\citenamefont {Page}(1993)}]{Page93Avg}%
  \BibitemOpen
  \bibfield  {author} {\bibinfo {author} {\bibfnamefont {D.~N.}\ \bibnamefont
  {Page}},\ }\bibfield  {title} {\bibinfo {title} {Average entropy of a
  subsystem},\ }\href {https://doi.org/10.1103/PhysRevLett.71.1291} {\bibfield
  {journal} {\bibinfo  {journal} {Phys. Rev. Lett.}\ }\textbf {\bibinfo
  {volume} {71}},\ \bibinfo {pages} {1291} (\bibinfo {year}
  {1993})}\BibitemShut {NoStop}%
\bibitem [{\citenamefont {Geraedts}\ \emph {et~al.}(2016)\citenamefont
  {Geraedts}, \citenamefont {Nandkishore},\ and\ \citenamefont
  {Regnault}}]{Geraedts2016}%
  \BibitemOpen
  \bibfield  {author} {\bibinfo {author} {\bibfnamefont {S.~D.}\ \bibnamefont
  {Geraedts}}, \bibinfo {author} {\bibfnamefont {R.}~\bibnamefont
  {Nandkishore}},\ and\ \bibinfo {author} {\bibfnamefont {N.}~\bibnamefont
  {Regnault}},\ }\bibfield  {title} {\bibinfo {title} {Many-body localization
  and thermalization: Insights from the entanglement spectrum},\ }\href
  {https://doi.org/10.1103/PhysRevB.93.174202} {\bibfield  {journal} {\bibinfo
  {journal} {Phys. Rev. B}\ }\textbf {\bibinfo {volume} {93}},\ \bibinfo
  {pages} {174202} (\bibinfo {year} {2016})}\BibitemShut {NoStop}%
\bibitem [{\citenamefont {Kj\"all}\ \emph {et~al.}(2014)\citenamefont
  {Kj\"all}, \citenamefont {Bardarson},\ and\ \citenamefont
  {Pollmann}}]{kjaell2014}%
  \BibitemOpen
  \bibfield  {author} {\bibinfo {author} {\bibfnamefont {J.~A.}\ \bibnamefont
  {Kj\"all}}, \bibinfo {author} {\bibfnamefont {J.~H.}\ \bibnamefont
  {Bardarson}},\ and\ \bibinfo {author} {\bibfnamefont {F.}~\bibnamefont
  {Pollmann}},\ }\bibfield  {title} {\bibinfo {title} {Many-body localization
  in a disordered quantum {Ising} chain},\ }\href
  {https://doi.org/10.1103/PhysRevLett.113.107204} {\bibfield  {journal}
  {\bibinfo  {journal} {Phys. Rev. Lett.}\ }\textbf {\bibinfo {volume} {113}},\
  \bibinfo {pages} {107204} (\bibinfo {year} {2014})}\BibitemShut {NoStop}%
\end{thebibliography}%

\end{document}